\documentclass[11pt]{article}
\usepackage[left=1.5cm, right=1.5cm, top=1.785cm, bottom=2.0cm]{geometry}
\usepackage{mathtools}
\usepackage{multicol}
\usepackage{subcaption}  
\usepackage{graphicx}
\usepackage{amsmath}
\usepackage{amssymb}
\usepackage{tikz}
\usetikzlibrary{arrows.meta, calc, decorations.markings, patterns.meta, shadings, decorations.pathmorphing}
\usetikzlibrary{external}
\usepackage{natbib}
\usepackage[noabbrev]{cleveref}

\definecolor{gel}{rgb}{0.82,0.84,0.92}
\definecolor{outline}{rgb}{0.18,0.44,0.73}

\newcommand{\dd}{\mathrm{d}}

\title{Hydrogel swelling in the osmosis-dominated regime}

\author{Ellen M. Jolley \and Thomas D. Montenegro-Johnson}

\begin{document}
\maketitle
\begin{center}
\small Mathematics Institute, Zeeman Building, University of Warwick, Coventry, CV4 7AL, UK \\
\end{center}
\begin{abstract}
    Hydrogels are soft, porous materials with the ability to absorb large amounts of water, swelling to many times their original volume. Modelling of hydrogel swelling is typically based on a free energy comprising a `mixing' (or osmotic) term and an elastic term. Here, we show that significant analytical reductions to the model can be made by exploiting the typically very small ratio between the characteristic scales of these osmotic and elastic energies, deriving a reduced set of governing equations which we term the `osmosis-dominated' regime. In the case of small deformation, we show that the osmosis-dominated equations can be reduced to a set of linear partial differential equations (a diffusion equation, a Poisson equation, and a Stokes-like set of equations) between which there is only one-way coupling, meaning they may be solved sequentially using classical techniques. In the nonlinear case, we propose new finite element formulations based on techniques from incompressible hyperelasticity. 
\end{abstract}

\section{Introduction}
\label{sec:introduction}
A hydrogel is a biphasic material comprising a fluid phase (water) and a solid phase consisting of a cross-linked network of hydrophilic polymers. Hydrogels may increase their volume several-fold by the absorption of water (swelling) \citep{tanaka1979kinetics}, and the water content of a swollen hydrogel may be as high as $99\%$ \citep{bertrand2016dynamics}. The swelling behaviour of hydrogels is often proposed as a useful actuation mechanism \citep{dong2021Programmable,cheng2021Recent,banerjee2018Hydrogel}, in particular for medical applications because being soft, relatively inert and having high water content makes them often biocompatible \citep{chirani2015history}. For example, hydrogel swelling has been proposed as a delivery mechanism for targeted drug therapies \citep{li2016designing}, a trigger for deployable medical devices such as gastrointestinal stents \citep{liu2023untethered}, or a gripper in soft robotics \citep{yuk2017Hydraulic}. 

The standard in hydrogel mechanics is based on Flory-Rehner theory \citep{flory1943statisticalA, flory1943statisticalB}, which posits a split in the free energy of the hydrogel into an `elastic' part (capturing the stretching of polymer chains) and a `mixing' part (capturing the chemical energy associated with mixing polymer and solvent). This essentially frames the phenomenon of swelling as a competition between elastic and chemical effects: from a chemical point of view, it is desirable to absorb water, but this results in an elastic penalty due to stretching polymer chains. Equilibrium is then attained when the two are in balance. With some notable exceptions, for example molecular dynamics and network theories \citep{lei2021Recent}, the bulk of hydrogel mechanics literature is built on this premise \citep{bertrand2016dynamics,webber2023linear,chester2011thermo,hong2008theory,quesada-perez2011Gel,urrea-quintero2024Comparative}. Among these are both equilibirum \citep{chester2011thermo,hong2008theory} and transient theories, which require an additional fluid transport law. Coupling a Flory-Rehner free energy to Darcy flow produces a poroelastic formulation \citep{biot1941general} with an additional `osmotic pressure' term capturing chemical effects \citep{doi2009gel,bertrand2016dynamics,webber2023linear}. 

A further distinguishing feature of these formulations is whether or not they assume small deformations, and thus whether their model results in linear or nonlinear governing equations. Using a linear system can enable analytic and semi-analytic solutions for hydrogel dynamics in simple geometries \citep{doi2009gel,jolley2026Asymptotica} and has been shown to agree well with experiments in the case of modest swelling ratios \citep{yamaue2004Theory,yoon2010Poroelastic}. Since any nonlinear constitutive law reduces to linear elasticity in the limit of small deformation, choosing a linear model also sidesteps the difficult issue of selecting an appropriate constitutive law faced by nonlinear formulations, which can have a significant impact on the output of a model \citep{urrea-quintero2024Comparative}. However, linear models are known to be inaccurate for larger deformations \citep{bouklas2012Swelling,macminn2016large}, which has led many authors to prefer a fully nonlinear formulation \citep{hong2008theory,chester2010Coupled,bertrand2016dynamics}. This comes at the expense of analytical simplicitly, and typically requires solution by finite element methods in all but the simplest of geometries. 

Several finite element formulations for large-deformation hydrogel mechanics have been developed, and a standard approach is yet to emerge \citep{urrea-quintero2024Comparative}. Many authors \citep{kang2010Variational,lucantonio2013Transient,bouklas2015Nonlinear,duan2013Simulation} build on an approach developed originally as an equilibrium solver by \citet{hong2009Inhomogeneous} and extended to the transient case by \citet{zhang2009Finite}, but the numerical implementation and constitutive details differ between authors. A related but distinct approach was developed by \citet{chester2015Finite}, which notably differs from \citet{hong2009Inhomogeneous} in its implementation of the incompressibility condition. There is a significant conceptual and technical hurdle involved in developing a bespoke hydrogel finite element solver, and it is therefore appealing to be able to analogise to a related problem with a more established suite of methods and convergence results available; for example, \citet{duan2013Simulation} achieve this for hydrogels by analogy with thermoelasticity. 

One option to achieve this that has been rarely explored in the context of hydrogel mechanics is the use of `staggered' or `split' finite element methods, which are common for poroelastic problems in geomechanics \citep{zienkiewicz1988Unconditionally}. For example, in the context of poroelasticity, a staggered method may allow the problem to be divided into separate solvers for diffusion (i.e. solvent transport) and solid mechanics; we provide more detail on the idea in \cref{sec:transient}. While staggered formulations are often less stable than their monolithic counterparts (i.e. those that solve transport and mechanics in a single step) \citep{kim2011Stabilitya}, they have the key advantage that they reduce the system into two well-studied forms for which numerous techniques, stability criteria, preconditioners and so on have already been developed. There have been few studies investigating the application of staggered techniques to hydrogel swelling problems. A review by \citet{urrea-quintero2024Comparative} found that staggered methods achieved comparable accuracy and stability to monolithic methods for a particular constitutive law (called `Constitutive Model II' in the paper) which is based on the formulation by \citet{chester2015Finite}. However, the authors noted the success of this method relied on a relation between pressure and porosity provided by the simple geometry they considered, leaving questions about the broader applicability of the method.

In this paper, we seek to investigate how the intricately coupled, nonlinear equations governing hydrogel mechanics can be simplified by exploiting realistic values of physical parameters; in particular, we claim that super-absorbent hydrogels generally exist within what we term the `osmosis-dominated' regime, wherein the characteristic scale associated with the mixing contribution to energy is significantly larger than that of the elastic contribution. This is corroborated by the values of these parameters predicted by polymer physics theory \citep{doi1995introduction} and experimental measurements \citep{horkay2010Structural,shao2023Independent,bertrand2016dynamics}. Indeed, such a separation of scales simply means that the energy penalty of stretching is significantly less than the energy benefit of absorbing water; thus if swelling is to be considered key to hydrogel behaviour, it could arguably be considered a defining property of a hydrogel. A recent experimental investigation by \citet{correas2026Hydrogel} makes the same observation and in fact, based on the results of their experiment, suggests some of the same modelling reductions which we derive independently and purely theoretically in this work. 

We will demonstrate that this regime enables a significant reduction in complexity in modelling both linear and nonlinear hydrogel mechanics. In the linear case, it is possible to reduce the hydrogel swelling problem into three hierarchically coupled linear PDE (partial differential equation) problems, resembling the diffusion equation, the Poisson equation and the Stokes equations respectively. This enables a systematic solution approach using classical techniques developed for these problems. In the nonlinear case, we show that equilibrium swelling problems in this regime are closely related to incompressible solid mechanics problems, and demonstrate that finite element techniques previously employed for incompressible hyperelasticity are applicable here. For transient nonlinear problems, we propose that the simplifications enabled by the regime enable a staggered finite element scheme in the manner of \citet{urrea-quintero2024Comparative} to be made more broadly applicable to hydrogel swelling problems (as opposed to only in simple geometries as suggested in that paper).

We begin from the framework of large-deformation poroelasticity as formulated by \citet{bertrand2016dynamics}. For clarity, we outline the full formulation in \cref{sec:framework}. In \cref{sec:osmosis}, we derive the osmosis-dominated regime which will form the basis for the remainder of the paper. In \cref{sec:linear}, we focus on small deformations, and show that the mechanics reduces to a hierarchy of linear partial differential equations that enable systematic solution procedure; as illustration, we solve the equations for a few example geometries. In \cref{sec:nonlinear}, we outline finite element formulations for solving osmosis-dominated hydrogel mechanics in both equilibrium and transient cases, utilising methods from incompressible hyperelasticity. In \cref{sec:conclusion}, we provide concluding remarks.

\section{Problem formulation and poroelastic hydrogel framework}
\label{sec:framework}
\begin{figure}
    \centering
    
            \tikzsetnextfilename{porous_domain}
        \begin{tikzpicture}
            \begin{scope}[shift={(-1.5,0)}]
                   \begin{scope}  
        \clip[preaction={shading=radial, left color=gel!80, right color=gel}, 
        ] plot [smooth cycle, tension=1] coordinates {
      (0:2)      
      (60:1.7)   
      (130:2.2)  
      (200:1.6)  
      (280:2.1)  
    };

        \foreach \i in {-2,...,2}{
            \foreach \j in {-2,...,2}{
            \pgfmathsetmacro{\x}{\i+rand*0.15};
            \pgfmathsetmacro{\y}{\j+rand*0.15};
            \pgfmathsetmacro{\r}{0.2+0.1*rand};
            \fill[outline!35] (\x,\y) circle (\r cm);
            \draw[outline!45] (\x,\y) circle (\r cm);
        }
        }
                   \end{scope}

            \draw[outline!90, line width=1.25mm] plot [smooth cycle, tension=1] coordinates {
      (0:2)
      (60:1.7) 
      (130:2.2) 
      (200:1.6)  
      (280:2.1)  
    };

\node[text width=3cm, align=center] at (0,0) {\LARGE $\Omega$};
\node[outline] at (2.3,-1.1) {\LARGE $\partial\Omega$};
    \end{scope}

    \begin{scope}[shift={(3,2)},scale=0.8]
        \begin{scope}
                \clip[preaction={shading=radial, left color=gel!80, right color=gel}, 
        ] (1,0) -- (4,0) to[out=45,in=270] (4.5,1) -- (0.5,1) to[out=270,in=135] (1,0); 

        \foreach \i in {0,...,4}{
            \foreach \j in {-2,...,2}{
            \pgfmathsetmacro{\x}{\i+rand*0.15};
            \pgfmathsetmacro{\y}{\j+rand*0.15};
            \pgfmathsetmacro{\r}{0.2+0.1*rand};
            \fill[outline!35] (\x,\y) circle (\r cm);
            \draw[outline!45] (\x,\y) circle (\r cm);
        }
        }
        \end{scope}
        \draw[thick] (0,0) -- (5,0);
        \foreach \i in {0,...,4}{
            \draw ({\i+0.5},0) -- ({\i+1},-0.3);
        }
                \draw[outline!50,line width=1.25mm] (1,0.1) to [out=135, in=270] (0.5,1);
                \draw[outline!50,line width=1.25mm] (4,0.1) to [out=45, in=270] (4.5,1);
        \draw[outline!90,line width=1.25mm,line cap=round] (1,0.1) -- (4,0.1);
        \draw[outline!50,thick,->] (4.5,0.5) -- (5,0.5);
        \draw[outline!50,thick,->] (0.5,0.5) -- (0,0.5);

        \node at (2.5,-0.7) {adhered rigid boundary};
    \end{scope}

        \begin{scope}[shift={(3,0)},scale=0.8]
        \begin{scope}
                \clip[preaction={shading=radial, left color=gel!80, right color=gel}, 
        ] (0.5,0) -- (4.5,0) -- (4.5,1) -- (0.5,1) -- (0.5,0); 

        \foreach \i in {0,...,4}{
            \foreach \j in {-2,...,2}{
            \pgfmathsetmacro{\x}{\i+rand*0.15};
            \pgfmathsetmacro{\y}{\j+rand*0.15};
            \pgfmathsetmacro{\r}{0.2+0.1*rand};
            \fill[outline!35] (\x,\y) circle (\r cm);
            \draw[outline!45] (\x,\y) circle (\r cm);
        }
        }
        \end{scope}
        \draw[thick] (0,0) -- (5,0);
        \foreach \i in {0,...,4}{
            \draw ({\i+0.5},0) -- ({\i+1},-0.3);
        }
                \draw[outline!50,line width=1.25mm] (0.5,0.1) -- (0.5,1);
                \draw[outline!50,line width=1.25mm] (4.5,0.1) -- (4.5,1);
        \draw[outline!90,line width=1.25mm,line cap=round] (0.5,0.1) -- (4.5,0.1);
                \draw[outline,thick,->] (4.65,0.25) -- (5.15,0.25);
        \draw[outline,thick,->] (0.35,0.25) -- (-0.15,0.25);
        \node at (2.5,-0.7) {rigid boundary with slip};
    \end{scope}

            \begin{scope}[shift={(3,-2)},scale=0.8]
        \begin{scope}
                \clip[preaction={shading=radial, left color=gel!80, right color=gel}, 
                ] (2.5,0) circle (1cm); 

        \foreach \i in {0,...,4}{
            \foreach \j in {-2,...,2}{
            \pgfmathsetmacro{\x}{\i+rand*0.15};
            \pgfmathsetmacro{\y}{\j+rand*0.15};
            \pgfmathsetmacro{\r}{0.2+0.1*rand};
            \fill[outline!35] (\x,\y) circle (\r cm);
            \draw[outline!45] (\x,\y) circle (\r cm);
        }
        }
        \end{scope}
        \draw[outline!90,line width=1.25mm] (2.5,0) circle (1cm);
        \foreach \i in {0,...,4}{
            \pgfmathsetmacro{\th}{\i*72};
            \draw[->,outline] ({2.5+1.2*cos(\th)},{1.2*sin(\th)}) -- ({2.5+1.5*cos(\th)},{1.5*sin(\th)});
        }
        \node at (2.5,-1.7) {free surface};
    \end{scope}
\end{tikzpicture}

    \caption{Left: A diagram showing the domain $\Omega$ of an arbitrarily shaped hydrogel and its boundary $\partial\Omega$. Right: examples of boundary conditions which may be applied on segments of $\partial\Omega$ according to relevant physical situations.}

\label{fig:domain}
\end{figure}
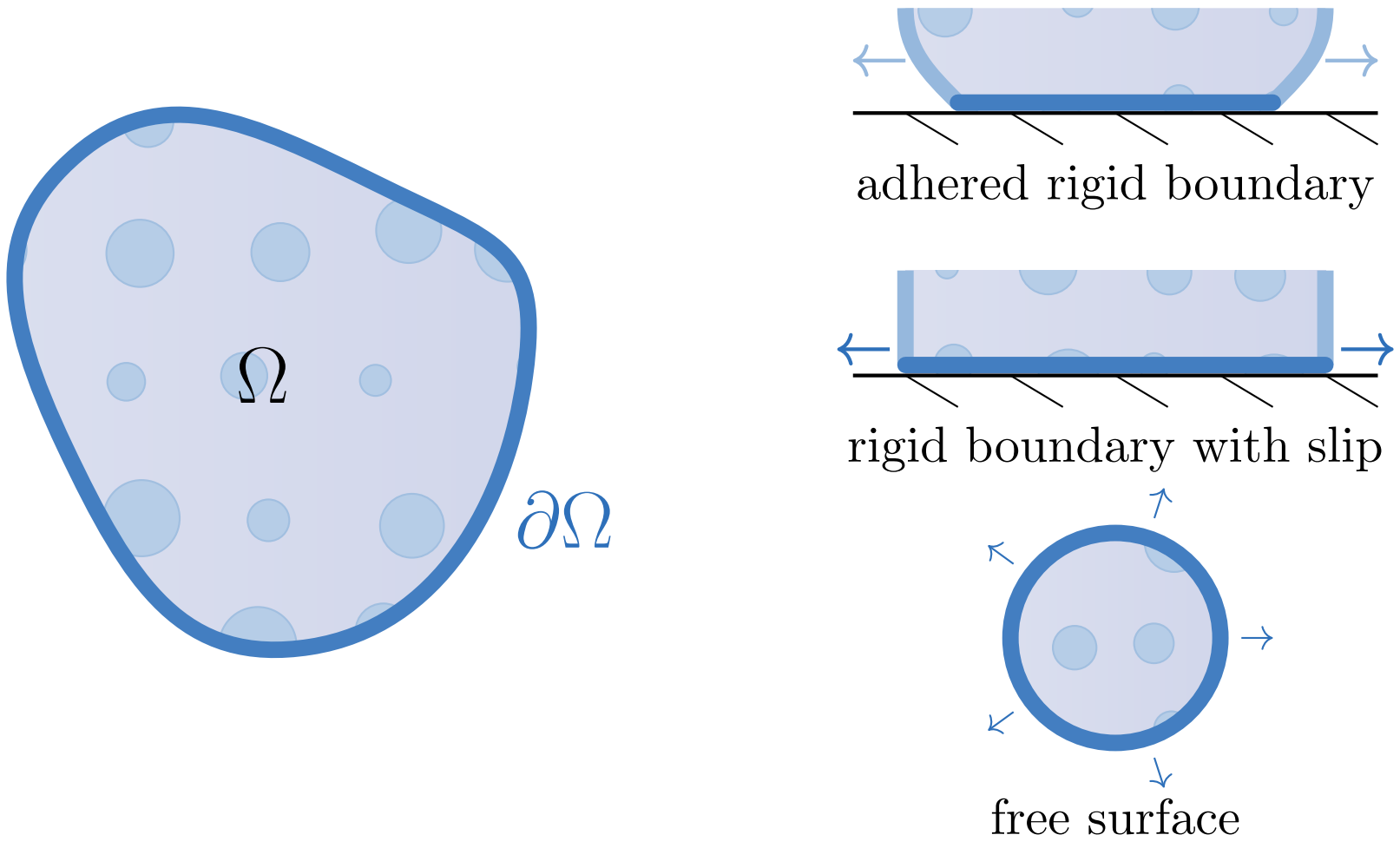

Consider a hydrogel which occupies a domain $\Omega$, with boundary $\partial\Omega$. On subsets of $\partial\Omega$, the gel may be free or somehow constrained; see \cref{fig:domain}. We detail the correspoinding boundary conditions in \cref{sec:boundaryconditions}.

We base our theory of hydrogel mechanics in the framework of large deformation poroelasticity, as set out by \citet{macminn2016large} and developed specifically for hydrogels by \citet{bertrand2016dynamics}; we summarize the full formulation below. In \cref{sec:osmosis}, we develop the reduced system termed the `osmosis-dominated regime' in this paper, which will form the basis for \cref{sec:linear} and \cref{sec:nonlinear}.

Throughout the paper, tilde ($\tilde{\ }$)  indicates a dimensional quantity and the absence thereof indicates a non-dimensional quantity. 
\subsection{Kinematics and two-phase formulation}
\label{sec:kinematics}
The gel is conceptualised as a two-phase mixture, consisting of a solid polymer network phase and a fluid phase, each of which is taken to be individually incompressible. The porosity $\phi$ reflects the local volume fraction of fluid in the mixture. We use an Eulerian formulation, meaning quantities are described with respect to the current configuration of the gel at time $\tilde{t}$. 

For nonlinear solid deformations, it is necessary to define a reference state for the solid phase of the gel. It is typical in solid mechanics to take the reference state to be the unstressed state of the material; however, it is difficult to determine where exactly this is for a hydrogel and authors have used different conventions in the literature. It is common to take the fully dry ($\phi = 0$) state as the unstressed reference state, however, as noted by \citet{webber2023linear}, the fully dry state is physically difficult to attain and it is unclear that a gel would be genuinely stress free in such a state. We therefore assume the reference state has some porosity $\phi_0$ such that $0 < \phi_0 < 1$. Taking the reference position of a material point which is at $\tilde{\mathbf{x}}$ at time $\tilde{t}$ to be $\tilde{\mathbf{x}}_0(\tilde{\mathbf{x}},\tilde{t})$, we characterise a deformation of the solid matrix by the displacement vector,
\begin{equation}
    \tilde{\boldsymbol{\xi}}(\tilde{\mathbf{x}},\tilde{t}) = \tilde{\mathbf{x}} - \tilde{\mathbf{x}}_0(\tilde{\mathbf{x}},\tilde{t}).
\end{equation}
We also define the (inverse) deformation gradient tensor,
\begin{equation}
    \mathbf{F}^{-1} = \tilde{\nabla}\tilde{\mathbf{x}}_0 = \mathbf{I}-\tilde{\nabla}\tilde{\boldsymbol{\xi}},
\end{equation}
and the Jacobian $J = \det(\mathbf{F})$. Here and throughout the paper, $\tilde{\nabla}$ denotes differentiation with respect to $\tilde{\mathbf{x}}$. We also use the convention that, in index notation, $F_{ij} = \partial \tilde{x}_i/\partial \tilde{x}_{0j}$. Since we assume both fluid and solid phases are incompressible, all volume change in the gel is related to uptake of water, and thus $J$ is directly related to $\phi$ by
\begin{equation}
    J =  \frac{1-\phi_0}{1-\phi}. \label{eq:jacobian}
\end{equation}
Mass conservation of the fluid and solid phases respectively dictate the balances,
\begin{equation}
    \frac{\partial\phi}{\partial \tilde{t}}+\tilde{\nabla}\cdot(\phi\tilde{\mathbf{v}}_f) = 0,
\label{eq:fluidcons} \end{equation} 
and
\begin{equation}
    \frac{\partial}{\partial \tilde{t}} (1-\phi)+\tilde{\nabla}\cdot[(1-\phi)\tilde{\mathbf{v}}_s] = 0, \label{eq:solidcons}
\end{equation}
where $\tilde{\mathbf{v}}_{f,s}$ are the fluid and solid velocities respectively. A standard result from the theory of nonlinear solid mechanics relates the solid velocity to the displacement by
\begin{equation}
    \tilde{\mathbf{v}}_s = \mathbf{F}\frac{\partial\tilde{\boldsymbol\xi}}{\partial \tilde{t}}. \label{eq:solidvel}
\end{equation}
Together, \cref{eq:fluidcons,eq:solidcons} imply
  \begin{equation} \tilde{\nabla}\cdot\tilde{\mathbf{q}} = 0,\label{eq:flux}
  \end{equation}
where $\tilde{\mathbf{q}} = \phi\tilde{\mathbf{v}_f}+(1-\phi)\tilde{\mathbf{v}}_s$ represents the volume-averaged material flux.

\subsection{Poromechanics}
\label{sec:poromechanics}
The interstitial fluid in the gel is assumed to be driven through the porous solid matrix by gradients in pore pressure $\tilde{p}$ according to Darcy's law,
\begin{equation}
    \phi(\tilde{\mathbf{v}}_f-\tilde{\mathbf{v}}_s) = -\frac{\tilde{k}(\phi)}{\tilde{\eta}_f}\tilde{\nabla}\tilde{p},\label{eq:darcy}
\end{equation}
where $\tilde{k}(\phi)$ is the permeability of the hydrogel, which depends on porosity, and $\tilde{\eta}_f$ is the dynamic viscosity of the fluid. Darcy's law is the leading-order homogenised transport law for Stokes flow through a rigid porous medium, so our use of it encodes the assumptions that the interstitial flow has negligible Reynolds number and is unaffected by the deformation of the solid matrix at leading order. 

In the gel mechanics literature, it is common to use an equivalent formulation of Darcy's law in terms of the chemical potential of the solvent by the substitution $\tilde{p} = \tilde{\mu}_f/\tilde{\Omega}_f$ \citep{bertrand2016dynamics,hong2008theory}. The reason for the equivalence is the assumption that fluid molecules have a constant volume $\tilde{\Omega}_f$ and that the fluid is incompressible; the energy change associated with the uptake of a fluid molecule $\tilde{\mu}_f$ is therefore directly related to the energy change associated with an increase in volume $\tilde{p}$. In this paper, we formulate our model in terms of pore pressure $\tilde{p}$.

Combining \cref{eq:fluidcons,eq:solidcons,eq:flux,eq:darcy}, we arrive at a nonlinear advection--diffusion equation for the porosity of the gel,
\begin{equation}
    \frac{\partial\phi}{\partial \tilde{t}} +\tilde{\mathbf{q}}\cdot\tilde{\nabla}\phi= \tilde{\nabla}\cdot\left[\frac{(1-\phi)\tilde{k}(\phi)}{\tilde{\eta}_f}\tilde{\nabla}\tilde{p}\right]. \label{eq:adv-diff}
\end{equation}
 
\subsection{Momentum balance}
\label{sec:momentum}
We use Flory-Rehner theory \citep{flory1943statisticalB} to construct the Helmholtz free energy density of the hydrogel. Flory-Rehner theory frames the hydrogel dynamics as being driven by a competition between elastic and chemical contributions; the chemical energy gained by absorbing water is countered by the energy expended in deforming the solid matrix. The free energy density per unit reference volume $\tilde{W}$ of the gel is then given by the sum of its elastic component $\tilde{W}_\text{elast}$ and the energy due to chemical mixing $\tilde{W}_\text{mix}$,
\begin{equation}
    \tilde{W} = \tilde{W}_{\text{elast}}+\tilde{W}_{\text{mix}}.
\end{equation}
The elastic energy of the gel is given as a function of the deformation by the constitutive law (we will consider both linear elastic and neo-Hookean constitutive laws in this work, as discussed in \cref{sec:constitutive}). The mixing contribution to free energy is derived from Flory-Huggins polymer theory (see \citet{doi1995introduction}),
\begin{equation}
    \tilde{W}_{\text{mix}} = \tilde{K}J\left[\phi\log\phi+\chi\phi(1-\phi)\right],
\end{equation}
where the constant $\tilde{K} = \tilde{k}_B\tilde{T}/\tilde{\Omega}_f$ is the characteristic osmotic energy of the gel (where $\tilde{k}_B$ is the Boltzmann constant, $\tilde{T}$ is the temperature and $\tilde{\Omega}_f$ is the volume of fluid per fluid molecule), and $\chi$ is the Flory mixing parameter, which quantifies the interaction energy between polymer and solvent. We take the mixing parameter to be a constant with the value 0.4 \citep{bertrand2016dynamics}. The stress tensor is then given by 
\begin{equation}
    \tilde{\boldsymbol\sigma} = - \tilde{p}\mathbf{I}+J^{-1}\frac{\partial\tilde{W}_\text{elast}}{\partial\mathbf{F}}\mathbf{F}^T +\frac{\partial\tilde{W}_\text{mix}}{\partial J}\mathbf{I}.\label{eq:stress}
\end{equation}
In the notation of Biot poroelasticity, this can be expressed as
\begin{equation}
     \tilde{\boldsymbol\sigma} = -(\tilde{p}+\tilde{\Pi})\mathbf{I}+\boldsymbol\sigma^\prime,
\end{equation}
where
\begin{equation}
    \tilde{\boldsymbol{\sigma}}^\prime = J^{-1}\frac{\partial\tilde{W}_{\text{elast}}}{\partial\mathbf{F}}\mathbf{F}^T,
\end{equation}
is the Terzaghi effective stress \citep{terzaghi1925erdbau}, and
\begin{equation}
    \tilde{\Pi}(\phi) = -\frac{\partial\tilde{W}_{\text{mix}}}{\partial J} = -\tilde{K}\left[\log\phi+\chi(1-\phi)^2+(1-\phi)\right],
\end{equation}
is the osmotic pressure, which is the additional thermodynamic pressure arising from the mixing free energy. Its effect in the model is to drive solvent towards regions with smaller $\phi$ (i.e. drier regions).

Conservation of momentum in the system is governed by Cauchy's momentum equation,
\begin{equation}
    \nabla\cdot\tilde{\boldsymbol\sigma} = \mathbf{0},
\end{equation}
or equiavlently,
\begin{equation}
    \tilde{\nabla} \tilde{p} +\nabla\tilde{\Pi}= \tilde{\nabla}\cdot\tilde{\boldsymbol\sigma}^\prime. \label{eq:cauchy}
\end{equation}
The gel dynamics are then primarily driven by the coupling between \cref{eq:cauchy} and \cref{eq:adv-diff}.

\subsection{Constitutive laws}
\label{sec:constitutive}
To close the system, it is necessary to specify a constitutive law (i.e. a relationship between $\tilde{W}$ and deformation) for the gel.  The constitutive law used in most nonlinear formulations of gel mechanics \citep{bertrand2016dynamics,hong2008theory,butler2022swelling,flory1943statisticalA} is the neo-Hookean constitutive law, which can be derived from an argument based on the statistical physics of Gaussian polymer chains. In this paper, we will use assume the gel is neo-Hookean throughout, although our methods can be applied to other constitutive laws. The strain energy density is given by
\begin{equation}
    \tilde{W}_{\text{elast}} = \frac{\tilde{\mu}_s}{2}\left[\mathrm{tr}\left(\mathbf{F}\mathbf{F}^T\right)-3-2\log J\right],
\end{equation}
where $\tilde{\mu}_s$ is the shear modulus of the gel. The effective stress tensor is then given by
\begin{equation}
    \tilde{\boldsymbol\sigma}^\prime = \tilde{\mu}_sJ^{-1}\left(\mathbf{F}\mathbf{F}^T-\mathbf{I}\right).
\end{equation}
In the case of small strain, expanding $\mathbf{F} = \mathbf{I}+\tilde{\nabla}\tilde{\boldsymbol\xi}$ for $|\tilde{\nabla}\tilde{\boldsymbol\xi}| \ll 1$ reduces the constitutive law to linear elastic,
\begin{equation}
    \tilde{\boldsymbol\sigma}^\prime = \tilde{\mu}_s\left[\tilde{\nabla}\tilde{\boldsymbol\xi}+\left(\tilde{\nabla}\tilde{\boldsymbol\xi}\right)^T\right].
\end{equation}
In this paper, we will consider both small strain linear reductions and fully nonlinear problems. 

\subsection{Non-dimensionalization}
\label{sec:nond}
We non-dimensionalize our equations according to
\begin{gather}
    (\tilde{\mathbf{x}},\tilde{\boldsymbol\xi},\tilde{\mathbf{x}}_0) = \tilde{L}(\mathbf{x},\boldsymbol{\xi},\mathbf{x}_0), \quad (\tilde{\boldsymbol\sigma}, \tilde{p}, \tilde{\Pi}) = \tilde{K}(\boldsymbol\sigma, p, \Pi), \\ \tilde{k} = \tilde{k}_0k,\quad \tilde{t} = \frac{\tilde{L}^2\tilde{\eta}_f}{\tilde{k}_0\tilde{K}}t, \quad (\tilde{\mathbf{v}}_{f,s},\tilde{\mathbf{q}}) = \frac{\tilde{k}_0\tilde{K}}{\tilde{\eta}_f\tilde{L}}(\mathbf{v}_{f,s},\mathbf{q}), \notag
\end{gather}
where $\tilde{L}$ is some measure of the size of the hydrogel, and $\tilde{k}_0$ is the permeability scale. To give some indication of typical values for the dimensional parameters used in our non-dimensionalization, \citet{bertrand2016dynamics} estimated $\tilde{\Omega}_f \approx 3\times 10^{-29}\ \mathrm{m}^3$ (which yields $\tilde{K} = 1.36\times 10^8\  \mathrm{Pa}$ at $\tilde{T} = 295\ \mathrm{K}$) and $\tilde{k}_0 = 8\times10^{-20}\ \mathrm{m}^2$. Given the viscosity of water is $\tilde{\eta}_f = 10^{-3}\ \mathrm{Pa}\ \mathrm{s}$, the result is a characteristic time scale for the gel swelling of between $10\ \mathrm{s}$ and $1000\ \mathrm{s}$ for a gel with size in the millimeter to centimeter range.  

The governing equations for the swelling of the gel are then given in non-dimensional form as follows. The momentum equation in the gel is 
\begin{equation}
    \nabla (p+\Pi)  = \nabla\cdot\boldsymbol\sigma', \label{eq:sys1}
\end{equation}
where, for convenience, we choose to shift the origin of osmotic pressure such that $\Pi(\phi_0) = 0$, i.e.
\begin{equation}
     \Pi(\phi) = -[\log\phi+\chi(1-\phi)^2+(1-\phi)]-\Pi_0,
\end{equation} 
where $\Pi_0 = -[\log\phi_0+\chi(1-\phi_0)^2+(1-\phi_0)]$. This is coupled to the constitutive law, 
\begin{equation}\label{eq:sys2}
    \boldsymbol\sigma' = GJ^{-1}(\mathbf{FF}^T-\mathbf{I}),
\end{equation}
where $G = \tilde{\mu}_s/\tilde{K} = \tilde{\Omega}_p/\tilde{\Omega_f}$ is a dimensionless number which captures the relative importance of elastic to chemical forces. The Jacobian and deformation gradient are related to displacement and porosity via
\begin{equation}
    \mathbf{F}^{-1} = \mathbf{I}-\nabla\boldsymbol\xi, \quad J = \det{\mathbf{F}} = \frac{1-\phi_0}{1-\phi}.\label{eq:sys3}
\end{equation}
This is coupled to the advection--diffusion equation for porosity,
\begin{equation}
    \frac{\partial{\phi}}{\partial t} +\mathbf{q}\cdot\nabla\phi = \nabla\cdot\left[(1-\phi)k(\phi)\nabla p\right],\label{eq:sys4}
\end{equation}
where
\begin{equation}
    \mathbf{q} = -k(\phi)\nabla p+\mathbf{v}_s, \quad \mathbf{v}_s = \mathbf{F}\cdot\frac{\partial\boldsymbol\xi}{\partial t}, \quad \nabla\cdot\mathbf{q} = 0.\label{eq:sys5}
\end{equation}

\subsection{Boundary conditions}
\label{sec:boundaryconditions}
On the boundary $\partial\Omega$, we impose a mechanical boundary condition and a permeation boundary condition \citep{doi2009gel}. The mechanical boundary condition depends on the geometric constraints of the problem and will generally entail prescribing stress, displacement or some combination on $\partial\Omega$ (or some subset of it; see \cref{fig:domain}). For example,
\begin{align}
    \boldsymbol\sigma\cdot\hat{\mathbf{n}} = \mathbf{t}_{\text{ext}}&\text{ on } \partial\Omega &\text{(free surface)},\label{eq:mech1}\\
     \text{or}\quad\xi = \xi_{\text{ext}}&\text{ on } \partial\Omega &\text{(fully adhered rigid surface)},\label{eq:mech2}
\end{align}
where $\mathbf{t}_\text{ext}$ represents the external traction vector, $\hat{\mathbf{n}}$ represents the unit outward normal to the boundary, and $\boldsymbol\xi_{\text{ext}}$ is some prescribed displacement. Note that in the case of a free surface the motion of the boundary must also be accounted for. The permeation condition expresses the degree to which solvent is able to permeate freely through the boundary, and may be of the below forms
\begin{align}
    p = p_\star &\text{ on } \partial\Omega&\text{(permeable boundary)},\label{eq:pperm}\\
    \text{or}\quad\hat{\mathbf{n}}\cdot\nabla p = 0  &\text{ on } \partial\Omega&\text{(impermeable boundary)},\label{eq:pimperm}
 \end{align}
where $p_\star$ is a constant pressure determined from the properties of the surrounding medium (in the case of water, it is the pressure of the surrounding water). A partially permeable boundary can also be captured by a linear combination of the two boundary conditions. 

We also require an initial condition on porosity $\phi$, which may be arbitrary in general. The initial displacement and pressure are determined by the initial porosity field and relevant boundary conditions.

\section{The osmosis-dominated limit}
\label{sec:osmosis}
The parameter termed $G$ in this paper captures the relative size of elastic energy contributions compared with osmotic energy contributions. Hydrogels, although a broad class of material with varying material properties, are typically soft and highly absorbent materials, which corresponds to a small $G$ value. As discussed in the introduction, both experimental measurements and theoretical predictions of this value tend to estimate it at around $10^{-4} - 10^{-3}$ \citep{cai2011mechanics,bertrand2016dynamics,correas2026Hydrogel}.

The focus of this paper is investigating the behaviour of the system \labelcref{eq:sys1,eq:sys2,eq:sys3,eq:sys4,eq:sys5} in the regime where $G$ is a small parameter, i.e. $G \ll 1$.  We will assume that as well as $G$ being small, that strain gradients remain $O(1)$, and thus that $G$ does indeed capture the characteristic scale of elastic forces compared with osmotic. In the case of sufficiently large strain gradients ($ \sim G^{-1}$), the two may be comparable despite the small $G$ value; the conclusions of this paper would not apply to such cases. 

Assuming strain gradients remain $O(1)$ and $G \ll 1$, the momentum balance \labelcref{eq:sys1} implies
\begin{equation}
    \nabla(p+\Pi) = O(G),
\end{equation}
and thus we expand
\begin{equation}
    p = p_0-\Pi(\phi) +GP, \label{eq:pexpand}
\end{equation}
where $P$ is an $O(1)$ function capturing the scaled total pressure in the gel. In other words, up to a constant, the pore pressure in the gel is equal to the negative osmotic pressure. The observation \labelcref{eq:pexpand} is not only a significant simplication to the modelling (since we now have the pore pressure as a function of porosity), but has in fact recently been supported by an experimental investigation \citep{correas2026Hydrogel} which also measured the pore pressure in a hydrogel to be equal to the negative osmotic pressure. In that work, the authors make the claim similar to that made here that this approximation holds regardless of applied loads and constitutive laws. The diffusion equation \labelcref{eq:sys4} at leading order is then
\begin{equation}
    \frac{\partial \phi}{\partial t}+\mathbf{q}\cdot\nabla\phi = -\nabla\cdot\left[(1-\phi)k(\phi)\nabla \Pi(\phi)\right]. \label{eq:diffeq}
\end{equation}
Note that the only explicit appearance of solid mechanics in \cref{eq:diffeq} is in the material flux $\mathbf{q}$. The permeation condition \labelcref{eq:pperm,eq:pimperm} now gives a condition on the porosity,
\begin{align}
    \phi &= \phi_\star &\text{(permeable boundary)},\label{eq:phiperm}\\
    \text{or}\quad\hat{\mathbf{n}}\cdot\nabla \phi &= 0 &\text{(impermeable boundary)}.\label{eq:phiimperm}
\end{align}
Thus \labelcref{eq:diffeq} is coupled to the solid deformation only via 
\begin{equation}
    \mathbf{q} = k(\phi)\nabla \Pi(\phi)+\mathbf{v}_s. \label{eq:qeq}
\end{equation}
The momentum equation \cref{eq:sys1} becomes
\begin{equation}
    \nabla P = \nabla \cdot\boldsymbol{\sigma}^\prime,\label{eq:momeq}
\end{equation}
where we have relabelled $\boldsymbol{\sigma}^\prime = J^{-1}(\mathbf{F}\mathbf{F}^T-\mathbf{I})$ (i.e. we have dropped the $G$ from \cref{eq:neohookean}). This is coupled to the porosity by the condition
\begin{equation}
    J = \frac{1-\phi_0}{1-\phi}.\label{eq:Jcond}
\end{equation}

The problem then takes the form of a diffusion problem coupled to a solid mechanics problem with the coupling occurring via \cref{eq:qeq,eq:Jcond} and, in the case of a free boundary, through the boundary conditions. We will see in \cref{sec:linear} that in the linear case this reduction enables significant analytical simplifications and in \cref{sec:nonlinear} that it enables simplified finite element computations through the use of established numerical routines as opposed to bespoke hydrogel solvers. 

\section{Linear poroelastic hydrogels in the osmosis-dominated regime}
\label{sec:linear}
In this section, we will focus on the small deformation limit of the osmosis-dominated regime we derived in \cref{sec:osmosis}. We therefore scale $\boldsymbol\xi = \delta\hat{\boldsymbol\xi}$. Since $\mathbf{F} = \mathbf{I}+\delta\nabla\hat{\boldsymbol{\xi}}$, a consequence of this is 
\begin{equation}
    J = \mathrm{det}\ \mathbf{F} = 1+\delta\mathrm{tr}\left(\nabla\hat{\boldsymbol{\xi}}\right)+O(\delta^2),
\end{equation}
and hence we expand
\begin{equation}
    \phi = \phi_0+\delta(1-\phi_0)\Phi,
\end{equation}
where $\Phi = \mathrm{tr}\left(\nabla\hat{\boldsymbol{\xi}}\right) = \nabla\cdot\hat{\boldsymbol{\xi}}$. Therefore the porosity does not vary substantially from its reference value. Since $\mathbf{q}$ is given by \cref{eq:qeq}, we find that the advection-diffusion equation \labelcref{eq:diffeq} reduces simply to a linear diffusion equation at leading order,
\begin{equation}
    \frac{\partial\Phi}{\partial t} = D\nabla^2\Phi,\label{eq:lindiff}
\end{equation}
with diffusion coefficient $D = -k(\phi_0)\Pi'(\phi_0)$. \Cref{eq:lindiff} is coupled to relevant permeation conditions \labelcref{eq:phiimperm,eq:phiperm} and the initial condition.

Note that because we assume small displacement, there is now no coupling from the solid mechanics to the diffusion equation at leading order, even in the case of a free boundary; the porosity evolves independently of the solid mechanics.

Turning to the solid mechanics problem, the momentum equation \labelcref{eq:momeq} becomes (at leading order)
\begin{equation}
        \nabla \hat{P} = \nabla\Phi+\nabla^2\hat{\boldsymbol{\xi}},\label{eq:swellStokes}
\end{equation}
where $\hat{P} = P/\delta$ is the scaled pressure and
\begin{equation}
    \Phi = \nabla\cdot\hat{\boldsymbol\xi} \label{eq:incomp}
\end{equation}
 is the known solution to the diffusion equation \labelcref{eq:lindiff}. This is coupled to the mechanical boundary condition, \labelcref{eq:mech1,eq:mech2}. Toegther, \cref{eq:lindiff,eq:swellStokes,eq:incomp}, form the fully linear gel mechanics model used by, for example, \citet{doi2009gel}.
 
 Decomposing $\hat{\boldsymbol\xi}$ into a solenoidal and irrotational part,
 \begin{equation}
    \hat{\boldsymbol{\xi}} = \nabla\alpha +\boldsymbol\xi^\prime\quad \text{with }\nabla\cdot\boldsymbol\xi^\prime = 0, \label{eq:helmholtz}
 \end{equation}
the problem may be divided into a Poisson problem for the quantity $\alpha$ (which we term `swelling potential') and a forced Stokes equation for the `incompressible correction' $\boldsymbol\xi^\prime$,
\begin{align}
     \nabla^2\alpha = \Phi, \label{eq:poisson}\\
    \nabla \hat{P} = \nabla^2\boldsymbol\xi^\prime+2\nabla\Phi,\quad\nabla\cdot\boldsymbol\xi^\prime = 0. \label{eq:Stokes}
\end{align}

The Poisson equation \labelcref{eq:poisson} requires either a Dirichlet condition on $\alpha$ (for a unique solution) or a Neumann condition (for uniqueness up to a constant). We have freedom in how we choose to assign this, since whatever boundary condition we impose on $\alpha$ can be adjusted for in the corresponding stress or displacement condition imposed on $\boldsymbol\xi^\prime$. In this paper, we will impose the Neumann condition
\begin{align}
   \hat{\mathbf{n}} \cdot\nabla\alpha = 0 &\text{ on }\partial\Omega. \label{eq:neumann}
\end{align}
This boundary condition has the advantage that it makes handling Dirichlet boundary conditions on displacement more straightforward; although we note that if any other boundary condition had proven more convenient we would be free to choose it provided we adjust the boundary conditions for the Stokes problem accordingly. Condition \labelcref{eq:neumann} only determines $\alpha$ up to a constant but this has no effect on the resulting physical fields. 

Thus our linear elastic hydrogel swelling problem in the osmotic-dominated limit is equivalent to a linear diffusion problem for the adjusted porosity $\Phi$, a Poisson problem \labelcref{eq:poisson} for the swelling potential $\alpha$, and a Stokes-like problem \labelcref{eq:Stokes} for the solenoidal part of displacement $\boldsymbol\xi^\prime$. Since many analytical and numerical solution techniques for such problems exist already, this formulation has the advantage that many hydrogel problems may then be solved by these same techniques. We illustrate the utility of this formulation for several such problems in \cref{sec:sphere,sec:arbitrary,sec:wedge} below.

\subsection{Swelling of a uniform sphere}
\label{sec:sphere}
\begin{figure}
    \centering
                       \tikzsetnextfilename{porous-circle}

        \begin{tikzpicture}
    
        \fill[shading=radial, left color=gel!80, right color=gel] (0,0) circle (2cm);
        \draw[outline!60,thick] (0,0) circle (2cm);
        \draw[outline!60,thick,dashed] (0,0) circle (2.2cm);
                    \begin{scope}
        \clip (0,0) circle (2cm);

        \foreach \i in {-2,...,2}{
            \foreach \j in {-2,...,2}{
            \pgfmathsetmacro{\x}{\i+rand*0.15};
            \pgfmathsetmacro{\y}{\j+rand*0.15};
            \pgfmathsetmacro{\r}{0.2+0.1*rand};
            \fill[outline!35] (\x,\y) circle (\r cm);
            \draw[outline!45] (\x,\y) circle (\r cm);
        }
        }
    \end{scope}

            \foreach \i in {0,...,11}{
            \pgfmathsetmacro{\th}{\i*30};
            \draw[->,outline!60] ({2*cos(\th)},{2*sin(\th)}) -- ({2.2*cos(\th)},{2.2*sin(\th)});
        }

\node[text width=3cm, align=center] at (0,0) {hydrogel \\ $\Phi = \Phi(r,t)$ \\ $\boldsymbol\xi = \left(\frac{d\alpha}{dr}+\xi^\prime_r\right)\mathbf{e}_r$};
 \node[anchor=north] at (1,-2) {$r=1$};
 \node[anchor=west,outline!70,align=left] at (2.3,0) {$\Phi=\Phi_\star$ \\ $\boldsymbol\sigma\cdot\hat{\mathbf{n}} = \mathbf{0}$};

            \begin{scope}[shift={(-4.5,2)}]
    \def\polarAngle{40}
    \def\radius{1.3} 

    \coordinate (rvec) at ({\radius*cos(\polarAngle)}, {\radius*sin(\polarAngle)});
    \draw[->, thick] (0,0) -- (rvec) node[anchor=south west] {$r$};

    \draw[->, thick] 
      ([shift={(0.7,0)}]0,0) arc[start angle=0, end angle=\polarAngle, radius=0.7];
    \node at (0.9,0.3) {$\theta$};
    \draw[dashed] (0,0) -- (\radius,0);
            \end{scope}
\end{tikzpicture}

    \caption{A diagram showing a hydrogel sphere with (linearized) porosity $\Phi$ and displacement $\boldsymbol\xi$, which is expressable in terms of the fields $\alpha$ and $\boldsymbol\xi^\prime$ \labelcref{eq:helmholtz}. We impose $\Phi = \Phi_\star$ and no normal stress on the free surface.}

\label{fig:sphere_diagram}
\end{figure}
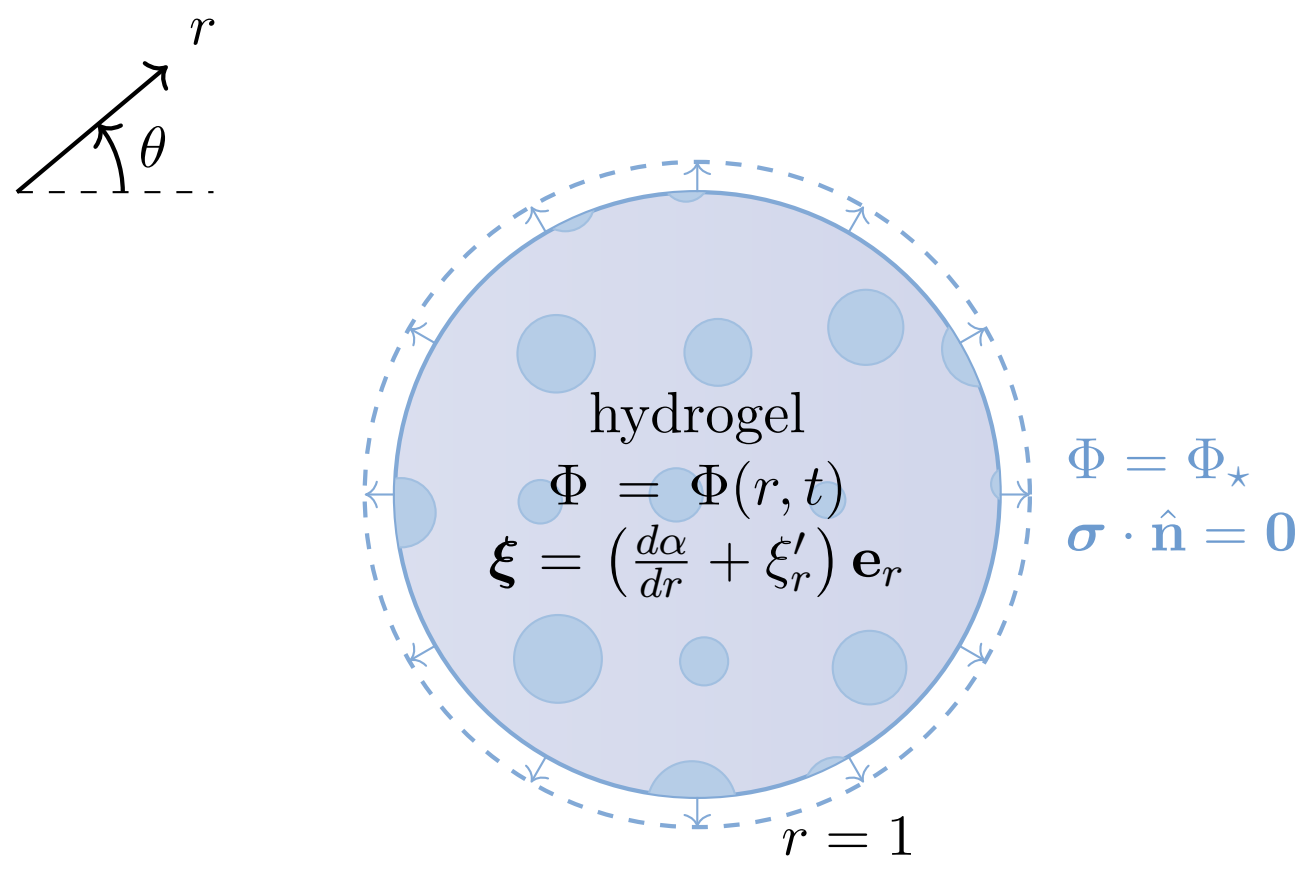

The swelling of a uniform sphere is often taken as the canonical hydrogel swelling problem, for example by \citet{bertrand2016dynamics,webber2023linear,butler2022swelling,punter2020Compressiona} among others. Being a fully isotropic, one-dimensional problem, it is reasonably straightforward to solve analytically in the linear case and numerically by finite differences in the nonlinear case. We highlight it here as a useful case study for understanding the mathematical structure set out in \cref{eq:helmholtz,eq:poisson,eq:Stokes}. We illustrate the problem in \cref{fig:sphere_diagram}.

We assume that the sphere is initially uniform with porosity at the reference value, i.e. $\Phi = 0$. The problem is spherically symmetric, so the equations set out in \cref{eq:lindiff,eq:incomp,eq:swellStokes,eq:helmholtz,eq:poisson,eq:Stokes,eq:neumann} then lead us to three linear PDE problems:

    \begin{enumerate}
            \item \hspace{0.1cm}\textbf{Diffusion problem}
            \begin{equation}
                \frac{\partial \Phi}{\partial t} = \frac{D}{r^2}\frac{\partial}{\partial r}\left(r^2\frac{\partial\Phi}{\partial r}\right);\quad\Phi(r,0) = 0;\quad\Phi(1,t) = \Phi_\star. \label{eq:diff-sphere}
            \end{equation}
           \item \hspace{0.1cm}\textbf{Poisson problem}
            \begin{equation}
                \frac{1}{r^2}\frac{\partial}{\partial r}\left(r^2\frac{\partial\alpha}{\partial r}\right) = \Phi;\quad \frac{\partial\alpha}{\partial r}(1,t) = 0.
                \label{eq:poisson-sphere}
            \end{equation}
      \item  \hspace{0.1cm}\textbf{Stokes-like problem}
        \begin{multline}
            \frac{\partial\hat{P}}{\partial r} = \frac{1}{r^2}\left[\frac{\partial}{\partial r}\left(\frac{r^2\partial\xi^\prime_r}{\partial r}\right)-2\xi^\prime_r\right]+2\frac{\partial\Phi}{\partial r};\quad
            \frac{1}{r^2}\frac{\partial}{\partial r}\left(r^2\xi^\prime_r\right) = 0; \\ -\hat{P}
            +2\frac{\partial\xi^\prime_r}{\partial r}
            = -2\frac{\partial^2\alpha}{\partial r^2}
        \text{ at } r=1.
            \label{eq:stokes-sphere}
        \end{multline}
    \end{enumerate}

The first of these, \cref{eq:diff-sphere}, is the classical problem of heating/cooling of a sphere \citep{logan1998Applied}. It admits a series solution which can be obtained by separation of variables,
\begin{equation}
    \Phi = \Phi_\star\left[1+\sum_{n\in \mathbb{N}} e^{-n^2\pi^2Dt}\frac{(-1)^n}{n\pi}\frac{\sin(n\pi r)}{r}\right]. \label{eq:phi-sphere}
\end{equation}
The Poisson problem can then be straightforwardly integrated to obtain
\begin{equation}
    \alpha = \Phi_\star\left[\frac{1}{6}r^2+2\sum_{n\in \mathbb{N}} e^{-n^2\pi^2Dt}\frac{(-1)^{n+1}}{n^3\pi^3}\frac{\sin(n\pi r)}{r}\right]+c_1(t),
\end{equation}
where $c_1(t) = \Phi_\star\left[1/3-2\sum_{n\in \mathbb{N}}e^{-n^2\pi^2Dt}/(n^2\pi^2)\right]$ is a function of time set by the boundary condition \labelcref{eq:neumann}; it has no impact on the resulting physical fields. 

\begin{figure}
    \centering
    \vspace{0.5cm}
    \begin{tabular}{c c c}
        & $\quad \Phi$ & $\quad \hat{\xi}_r$ \\
        $t=0.01$ & \begin{minipage}{0.3\textwidth}
            \includegraphics[width=\textwidth]{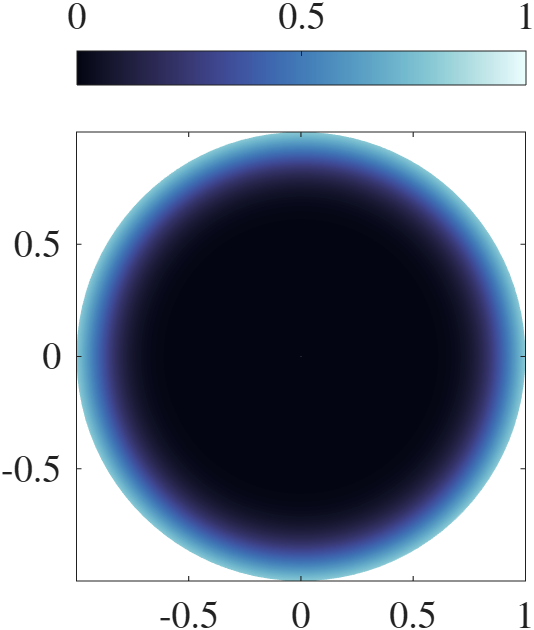}
        \end{minipage} &
        \begin{minipage}{0.3\textwidth}
            \includegraphics[width=\textwidth]{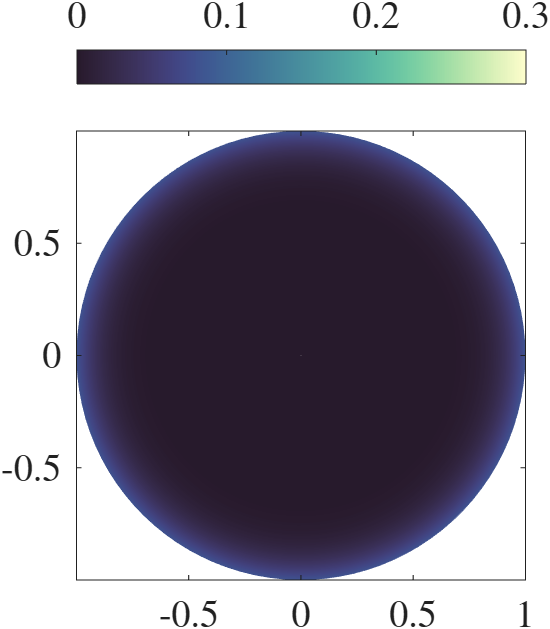}
        \end{minipage} \\
        $t=0.1$ &\begin{minipage}{0.3\textwidth}
            \includegraphics[width=\textwidth,trim=0 0 0 1.5cm,clip]{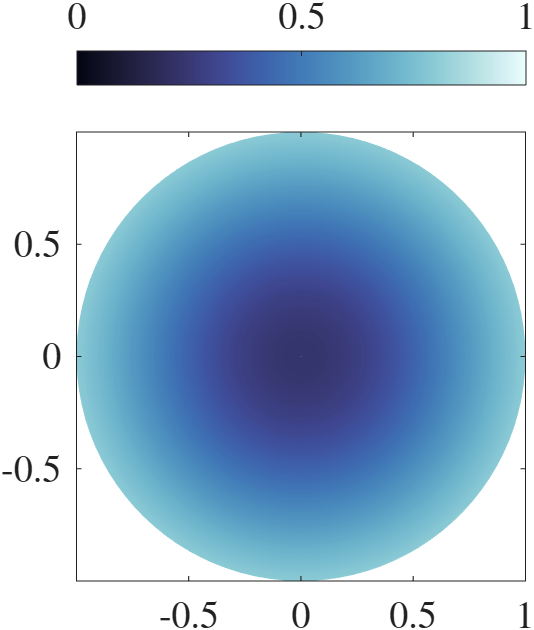}
        \end{minipage} &
        \begin{minipage}{0.3\textwidth}
            \includegraphics[width=\textwidth,trim=0 0 0 1.5cm,clip]{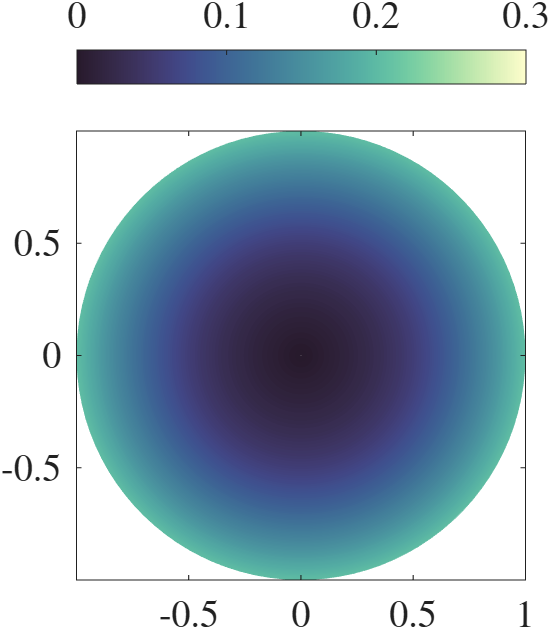}
        \end{minipage} \\
        $t = 0.5$ & \begin{minipage}{0.3\textwidth}
            \includegraphics[width=\textwidth,trim=0 0 0 1.5cm,clip]{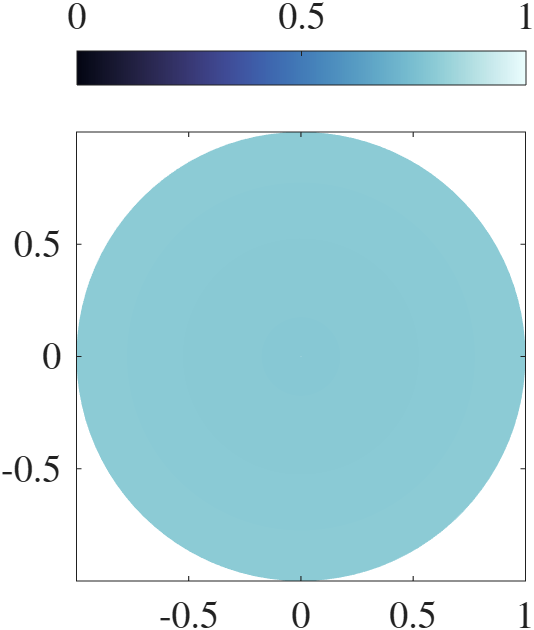}
        \end{minipage} &
        \begin{minipage}{0.3\textwidth}
            \includegraphics[width=\textwidth,trim=0 0 0 1.5cm,clip]{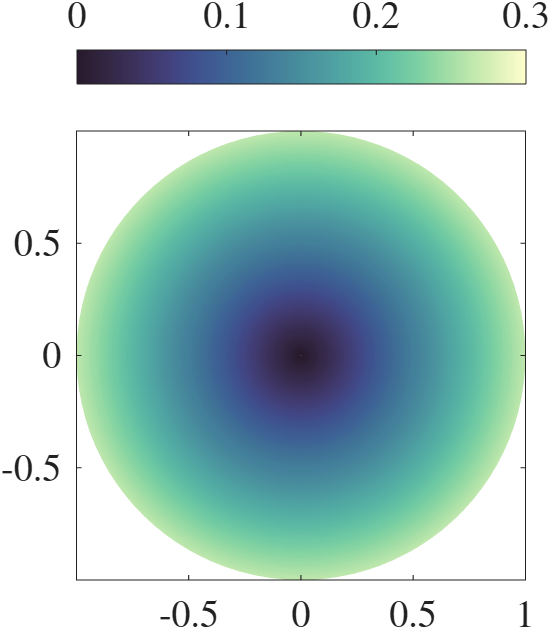}
        \end{minipage} \\
    \end{tabular}
    \caption{Plots of the solutions \labelcref{eq:phi-sphere,eq:xi-sphere} for the problem set out in \cref{sec:sphere}, at times $t = 0.01,\ 0.1,\ 0.5$, with $\Phi$ on the left and $\hat{\xi}_r = \hat{\boldsymbol{\xi}}\cdot\mathbf{e}_r$ on the right. The $t = 0.5$ case has effectively converged to the steady state. Parameters were $D = 1$ and $\Phi_\star = 0.8$. The radius of the sphere only changes at $O(\delta)$ so we have neglected this in our depiction.}
    \label{fig:sphere_results}
\end{figure}

Finally, the Stokes problem \labelcref{eq:stokes-sphere} has solution
\begin{equation}
    \xi^\prime_r = 0, \quad \hat{P} = -2\Phi+c_2(t),
\end{equation}
where $c_2(t) =  \Phi_\star\left[1/3+2\sum_{n\in \mathbb{N}}e^{-n^2\pi^2Dt}/(n^2\pi^2)\right]$ is set by the stress boundary condition. The final displacement field is therefore given by
\begin{equation}
    \hat{\boldsymbol\xi} = \nabla\alpha = \Phi_\star\left[\frac{1}{3}r+2\sum_{n\in \mathbb{N}} e^{-n^2\pi^2Dt}\frac{(-1)^{n+1}}{n^3\pi^3}\left(\frac{n\pi\cos(n\pi r)}{r}-\frac{\sin(n\pi r)}{r^2}\right)\right]\mathbf{e}_r. \label{eq:xi-sphere}
\end{equation}
We display the results in \cref{fig:sphere_results}. We see the scaled porosity $\Phi$ evolves from a uniform value of 0 (with a discontinuity at the boundary)
 to a uniform value of $\Phi_\star$, while the displacement converges to the non-uniform state predicted by \cref{eq:xi-sphere}.

\subsection{Response to arbitrary initial conditions}
\label{sec:arbitrary}
The solution of the diffusion equation in free space with a point source as an initial condition is known as the fundamental solution since the solution to the problem with arbitrary initial condition can be constructed from it \citep{logan1998Applied}. In our linear reduction, it is straightforward to calculate the displacement associated with a porosity given by the fundamental solution, and in this way determine hydrogel behaviour in response to an arbitrary initial porosity field.

For simplicity we consider here the case of symmetry in the third dimension, i.e. we treat the hydrogel as planar (in the sense of plane strain), $\Omega = \mathbb{R}^2$. We define split our problem into three linear PDE problems as follows:

    \begin{enumerate}
            \item \hspace{0.1cm}\textbf{Diffusion problem}
            \begin{equation}
                \frac{\partial \Phi}{\partial t} = D\nabla^2\Phi;\quad\Phi(\mathbf{x},0) = \Phi_I(\mathbf{x});\quad\Phi \to 0 \text{ as } |\mathbf{x}| \to \infty. \label{eq:diff-free}
            \end{equation}
           \item \hspace{0.1cm}\textbf{Poisson problem}
            \begin{equation}
                \nabla^2\alpha = \Phi;\quad |\nabla\alpha| \to 0 \text{ as }|\mathbf{x}| \to \infty.
                \label{eq:poisson-free}
            \end{equation}
      \item  \hspace{0.1cm}\textbf{Stokes-like problem}
        \begin{equation}
            \nabla\hat{P} = 2\nabla\Phi+\nabla^2\boldsymbol\xi^{\prime};\quad
            \nabla\cdot\boldsymbol\xi^\prime = 0;\quad |\boldsymbol\xi^\prime| \to 0 \text{ as } |\mathbf{x}|\to \infty.
            \label{eq:stokes-free}
        \end{equation}
    \end{enumerate}
Here $\mathbf{x}$ is a two-dimensional spatial coordinate.

\begin{figure}
    \centering
    \vspace{0.5cm}
    \begin{tabular}{c c}
         $\quad \Phi_I$ & $\quad \hat{\boldsymbol\xi}$ \\
         \begin{minipage}{0.3\textwidth}
            \includegraphics[width=\textwidth]{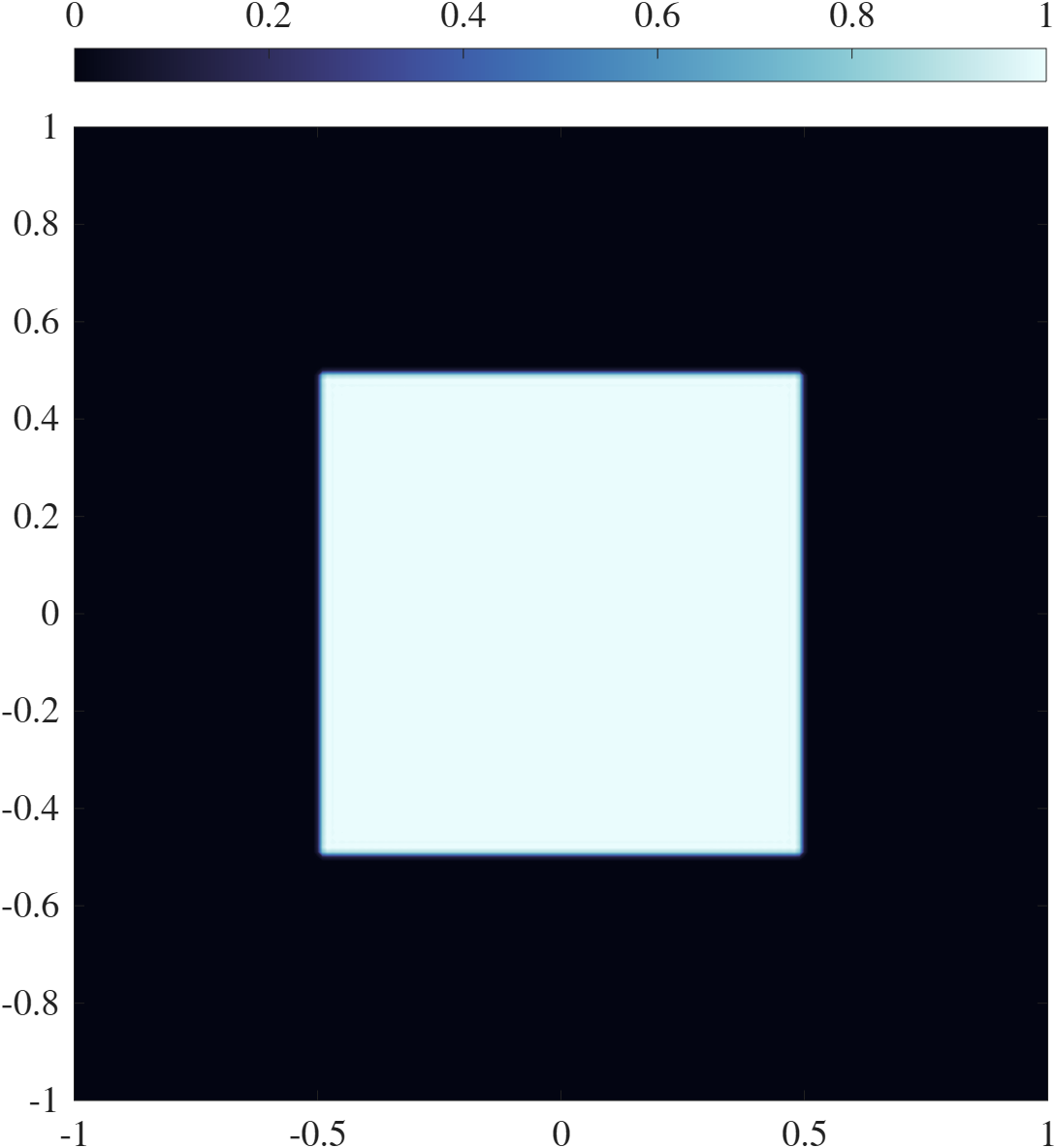}
        \end{minipage} &
        \begin{minipage}{0.3\textwidth}
            \includegraphics[width=\textwidth]{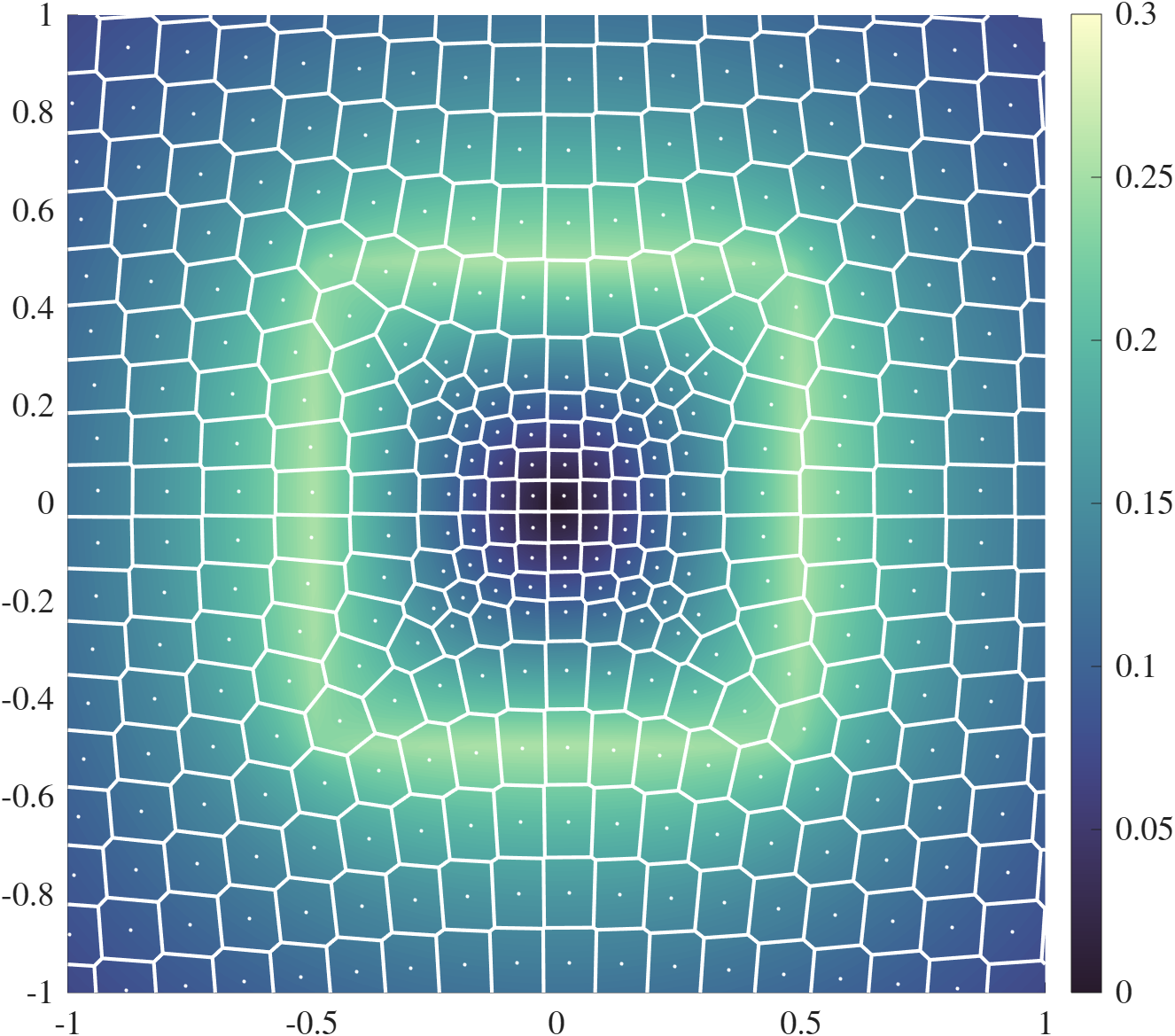}
        \end{minipage} \\
        \begin{minipage}{0.3\textwidth}
            \includegraphics[width=\textwidth,trim=0 0 0 1.5cm,clip]{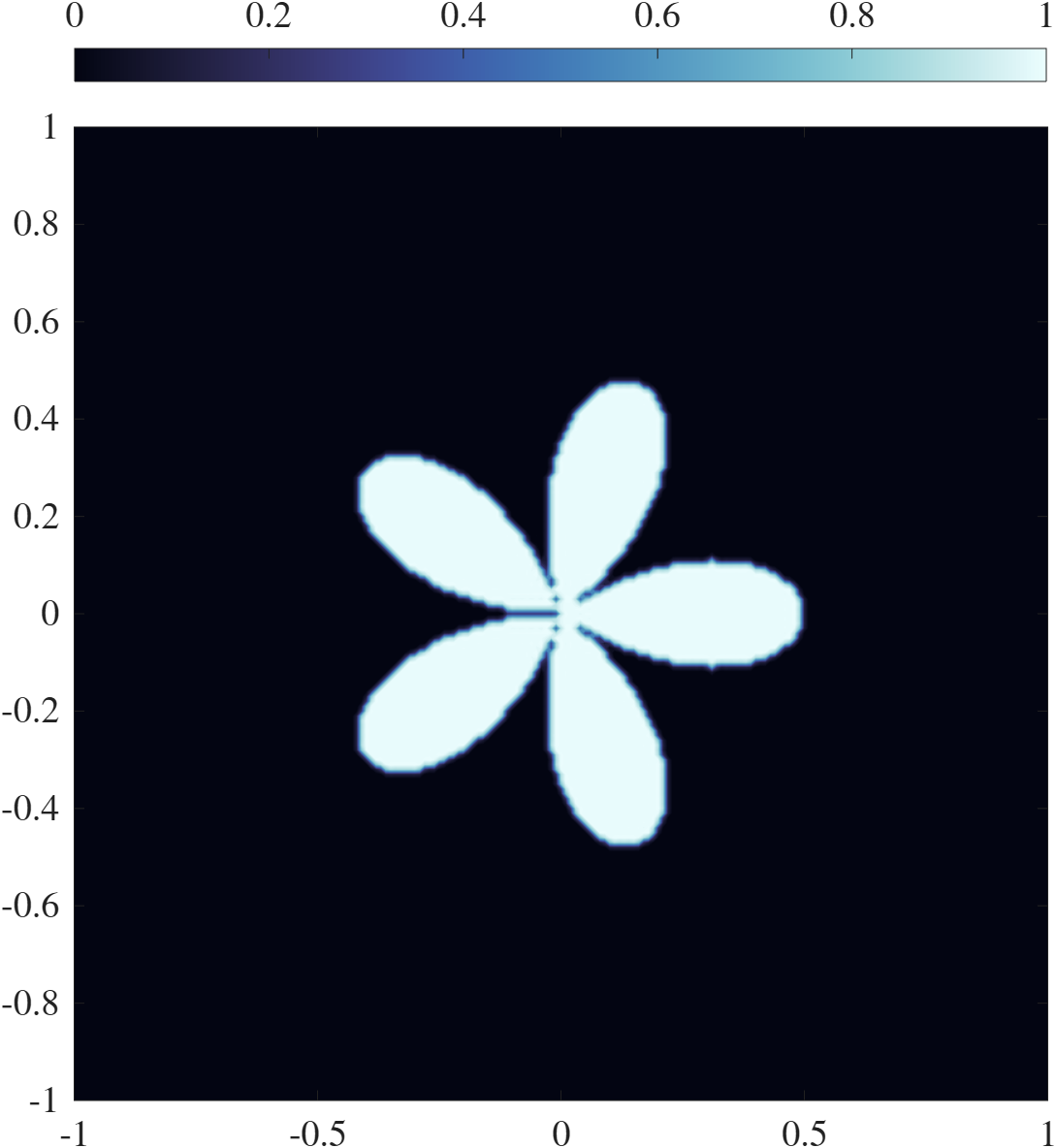}
        \end{minipage} &
        \begin{minipage}{0.3\textwidth}
            \includegraphics[width=\textwidth]{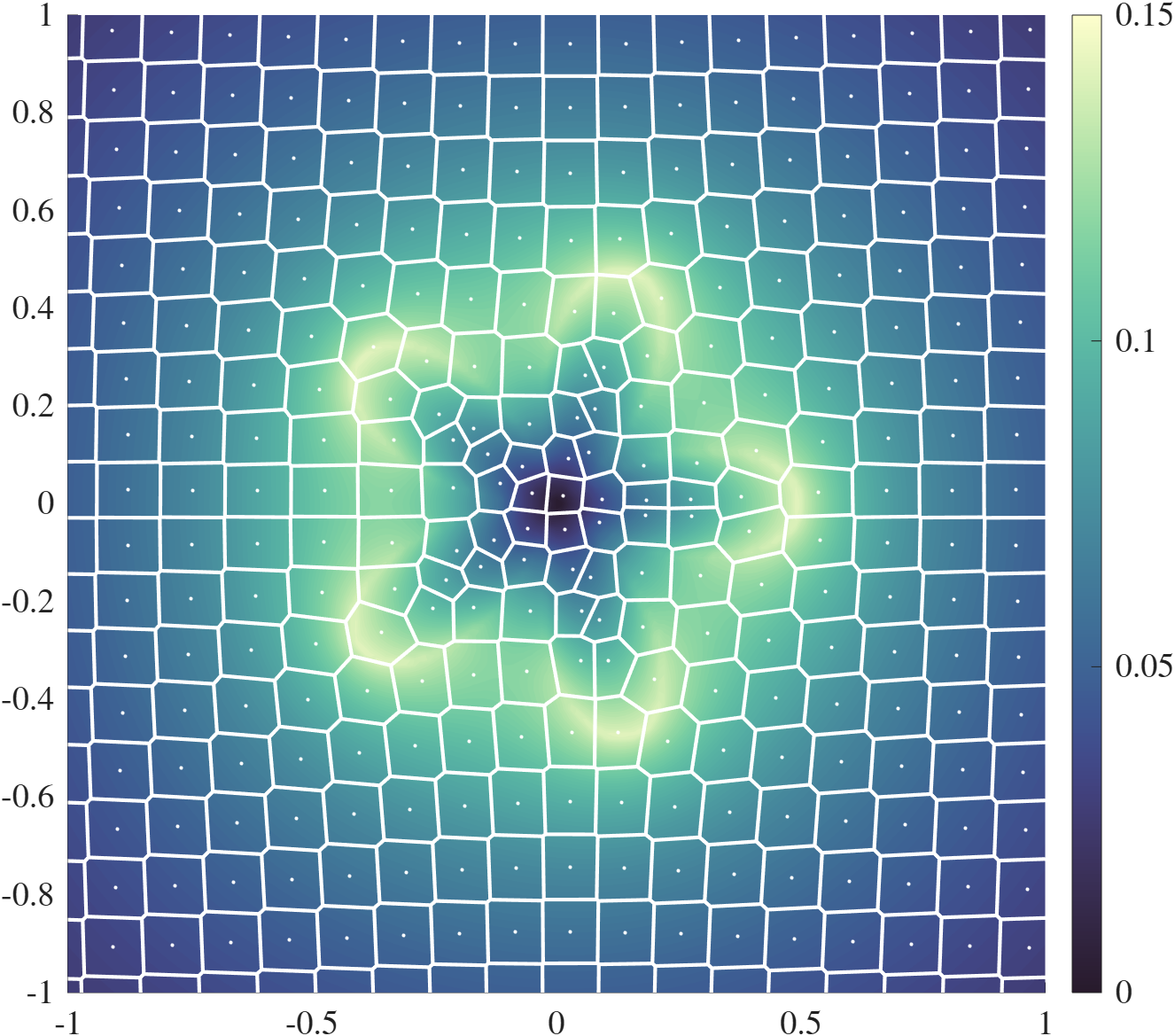}
        \end{minipage} \\
        \begin{minipage}{0.3\textwidth}
            \includegraphics[width=\textwidth,trim=0 0 0 1.5cm,clip]{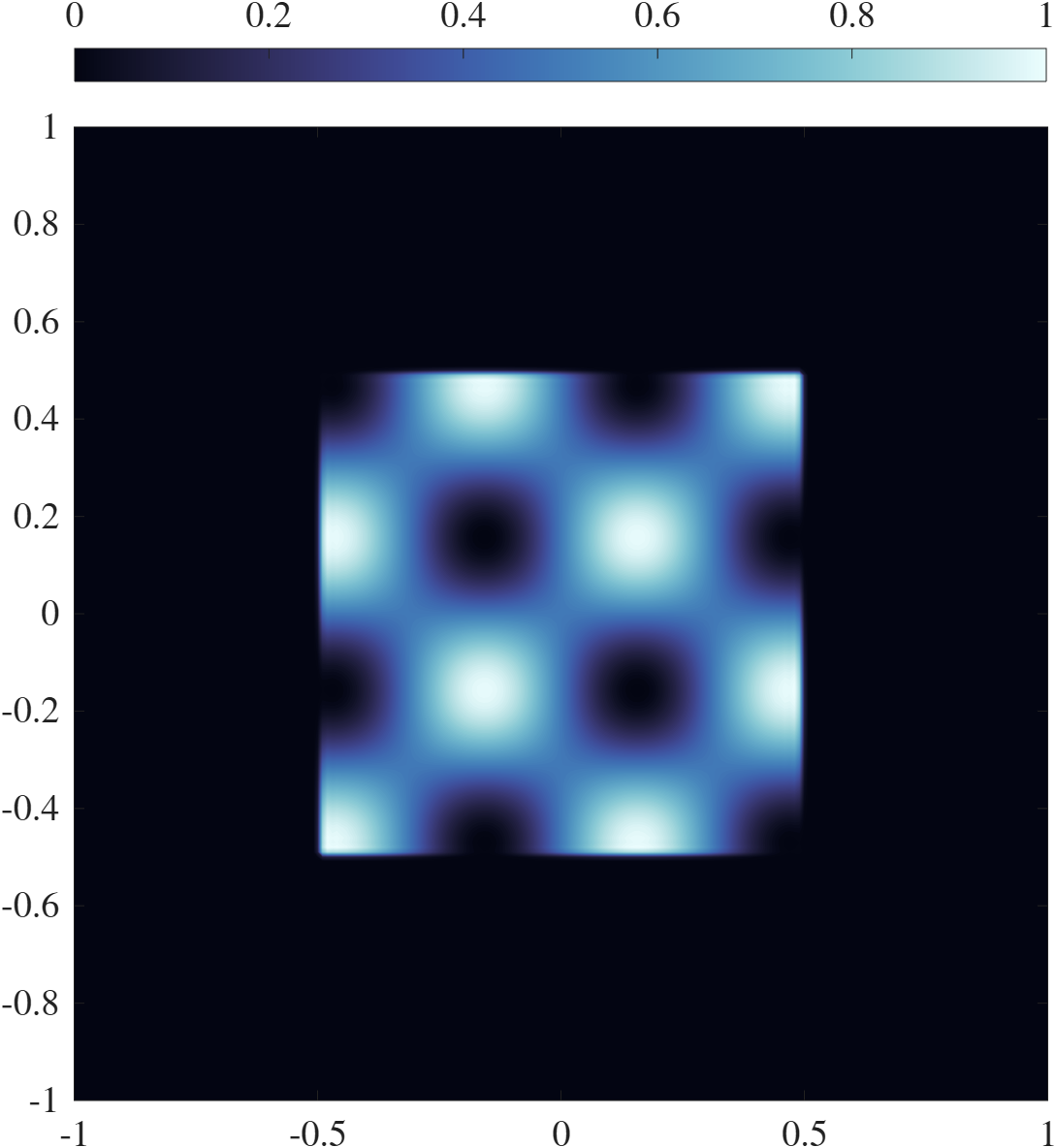}
        \end{minipage} &
        \begin{minipage}{0.3\textwidth}
            \includegraphics[width=\textwidth]{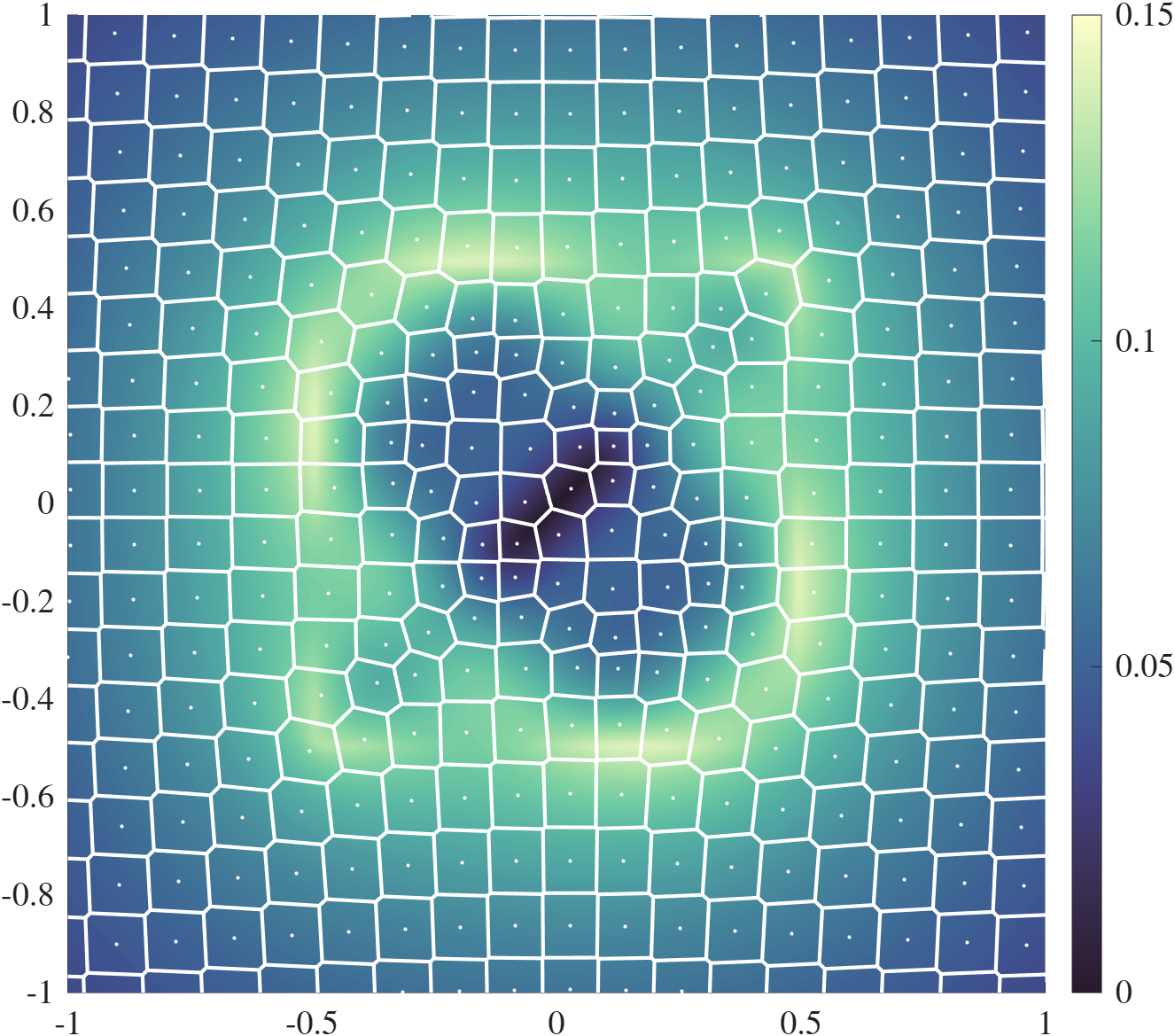}
        \end{minipage} 
    \end{tabular}
    \caption{Left: Initial conditions $\Phi_I$ for the problem \labelcref{eq:diff-free}. Right: Displacement of points at the end state (i.e. $t \to \infty$) for the corresponding initial condition on the left. The Voronoi tesselation shows the motion of a set of sample material points which were in a uniform grid at $t =0$ while the colour displays the magnitude of displacement.}
    \label{fig:free}
\end{figure}

The solutions to these problems \labelcref{eq:stokes-free,eq:diff-free,eq:poisson-free} are easily obtainable as
\begin{align}
    \Phi &= \Phi_I \ast G_1, \\
    \alpha &= \Phi \ast G_2,\\
    \boldsymbol\xi^\prime &= \mathbf{0},
\end{align}
where
\begin{equation}
    G_1(\mathbf{x}) = \frac{1}{4\pi D t}\exp\left(-\frac{|\mathbf{x}|^2}{4Dt}\right),\quad G_2(\mathbf{x}) = -\frac{1}{2\pi}\log |\mathbf{x}|,
\end{equation}
are the Green's functions for the heat equation \cref{eq:diff-free} and Poisson equation \cref{eq:poisson-free} respectively, and $\ast$ represents convolution. We illustrate some representative solutions to this problem in \cref{fig:free}, showing the equilibrium swelling/shrinking patterns produced by different initial porosity distributions. We see in general the largest displacement corresonds to regions where a large gradient in porosity was initially present.  

\subsection{Swelling of an adhered wedge}
\label{sec:wedge}
An advantage of reducing the hydrogel equations to this more recognisable form is that any geometries in which the Stokes equations are known to have analytic solutions are likely to also admit analytic solutions for hydrogel swelling. As an example, we consider a hydrogel swelling in an infinite wedge \citep{moffatt1964Viscousa}. While a truly infinite wedge is unlikely to arise in practice, we note that as with the equivalent Stokes solution it can be expected to approximate the behaviour of a hydrogel in the vicinity of a corner more generally.

Assuming plain strain and working in polar coordinates $(r,\theta)$, we consider a wedge-shaped hydrogel with fully adhered to rigid walls at $\theta = \pm \beta$ where $0 < \beta < \pi/2$. We seek equilibrium solutions for which
\begin{equation}
    \Phi = \Phi_\star,
\end{equation}
for some constant $\Phi_\star$. The solid deformation problem may then be divided into:
\begin{enumerate}
               \item \hspace{0.1cm}\textbf{Poisson problem}
            \begin{equation}
                \nabla^2\alpha = \Phi_\star;\quad \partial\alpha/\partial\theta = 0 \text{ at } \theta = \pm \beta.
                \label{eq:poisson-wedge}
            \end{equation}
      \item  \hspace{0.1cm}\textbf{Stokes-like problem}
        \begin{equation}
            \nabla\hat{P} = \nabla^2\boldsymbol\xi^{\prime};\quad
            \nabla\cdot\boldsymbol\xi^\prime = 0;\quad \boldsymbol{\xi}^\prime =  -\nabla\alpha \text{ at } \theta = \pm \beta.
            \label{eq:stokes-wedge}
        \end{equation}
\end{enumerate}

\begin{figure}
    \centering
    \begin{tabular}{c c c c}
        $\beta = \pi/10$ & $\beta = \pi/4$ & $\beta = 5\pi/12$ & \\
    \begin{minipage}{0.3\textwidth}
        \centering
        \includegraphics[width=0.85\textwidth,trim=0 0 1.7cm 0,clip]{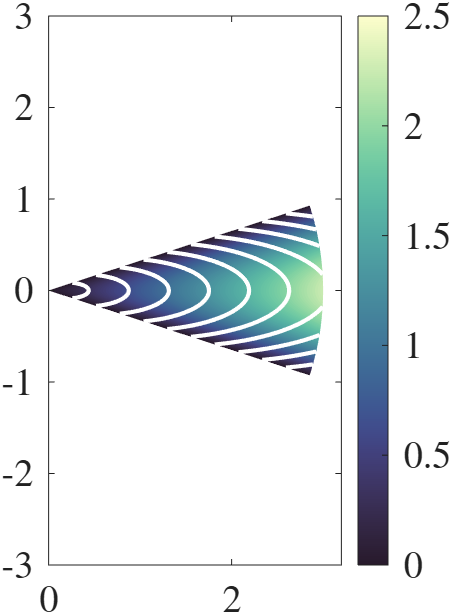}
    \end{minipage}&
    \begin{minipage}{0.3\textwidth}
        \centering
        \includegraphics[width=0.85\textwidth,trim=0 0 1.7cm 0,clip]{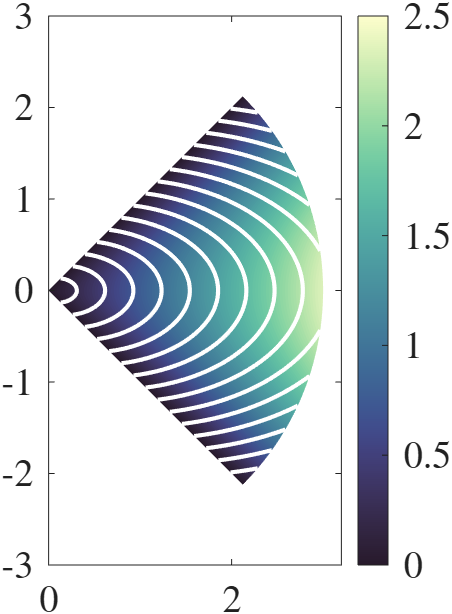}
    \end{minipage}&
    \begin{minipage}{0.3\textwidth}
        \includegraphics[width=\textwidth]{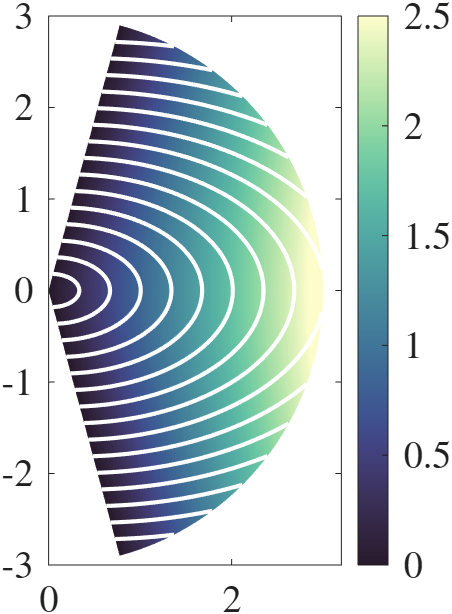}
    \end{minipage}&
        \begin{minipage}{0.08\textwidth}
        $|\hat{\boldsymbol\xi}|$
    \end{minipage}
\end{tabular}
    \caption{The solution \labelcref{eq:wedge-soln} for three values of $\beta$. The contours display material curves which were exact circular arcs in the reference configuration, i.e. $|\mathbf{x}_0| = \text{const}$. Note the wedge is infinite so the circular arc is simply the point at which the visualization is cut off, rather than a free surface. Parameters were $\Phi_\star = 1$, and $\delta = 0.5$.}
    \label{fig:wedge}
\end{figure}

\Cref{eq:poisson-wedge} integrates straightforwardly, yielding
\begin{equation}
    \alpha = \frac{1}{4}\Phi_\star r^2.
\end{equation}
The Stokes-like problem \labelcref{eq:stokes-wedge} can be solved by seeking a solution with streamfunction of form $\psi = \frac{1}{2}\Phi_\star r^2 g(\theta)$ for some function $g$ to be determined. The biharmonic equation yields an ODE for $g$,
\begin{equation}
    \frac{\dd^4g}{\dd\theta^4}+4\frac{\dd^2g}{\dd\theta^2} = 0;\quad g(\pm \beta) = 0,\quad \frac{\dd g}{\dd\theta}(\pm \beta) = -1.
\end{equation}
This is solved by $g(\theta) = \frac{1}{2}\left(\frac{\cos 2\theta-\cos 2\beta}{\sin 2\beta}\right)$, which yields the displacement field
\begin{equation}
    \hat{\boldsymbol\xi} = \frac{1}{2}\Phi_\star r \left(1-\frac{\sin 2 \theta}{\sin 2\beta}\right)\mathbf{e}_r-\frac{1}{2}\Phi_\star r \left(\frac{\cos 2\theta-\cos 2\beta}{\sin 2\beta}\right)\mathbf{e}_\theta. \label{eq:wedge-soln}
\end{equation}
We plot the solution for three values of $\beta$ in \cref{fig:wedge}. We see that initially circular material curves become elongated as the gel swells, and this effect is more pronounced at smaller angles.

\section{Nonlinear osmosis-dominated hydrogel swelling by finite element methods}
\label{sec:nonlinear}
In this section, we will consider methods of solution for the fully nonlinear system of osmosis-dominated equation, \labelcref{eq:pexpand,eq:diffeq,eq:qeq,eq:momeq,eq:Jcond}. By fully nonlinear, we mean that we include the nonlinear effects of the osmotic pressure, the constitutive law and the motion of the solid, and of any free surfaces in the problem.

In a fully nonlinear setup, the diffusion problem \labelcref{eq:diffeq} does not fully decouple from the solid problem \labelcref{eq:momeq}: the advection term $\mathbf{q}$ is related to the solid velocity and furthermore from a Lagrangian point of view the spatial derivatives also depend on deformation.  However, there remains a substantial simplification from the system with $G = O(1)$; namely,
\begin{itemize}
    \item the elastic stress which would otherwise appear in the diffusion equation is absent,
    \item the solid problem is reduced to a form closely related to incompressible hyperelasticity, with modified mass conservation criterion \labelcref{eq:Jcond},
    \item the boundary conditions \labelcref{eq:phiperm,eq:phiimperm} can now be expressed as boundary conditions on $\phi$ rather than on $p$ and on stress.
\end{itemize}
These factors have several advantages in the implementation of numerical methods. Firstly, though not eliminated, the coupling between the processes is weakened. The form of the equations is also more similar to the very well-studied systems of incompressible hyperelasticity and diffusion, making it easier to apply methods and insights originally developed for these problems. It is also easier to use $\phi$ as an independent variable as a result of the simplified boundary conditions; many other formulations use chemical potential (comparable to $p$ in our model) as the independent variable \citep{hong2009Inhomogeneous,kang2010Variational,chester2015Finite}.

The solid problem, as in the linear case, depends on porosity via the mass conservation criterion \labelcref{eq:Jcond}. The form of the equations \labelcref{eq:momeq,eq:Jcond} is therefore similar to a standard solid deformation problem for an incompressible neo-Hookean material but with a modified incompressible criterion to account for volumetric swelling. It is natural then to imagine that similar techniques to those used for incompressible nonlinear solid mechanics may be applicable to the present hydrogel swelling problem. Solutions to such problems in all but the simplest of cases are only available numerically, typically by finite element methods. To avoid numerical issues such as volumetric locking, a mixed $(u,p)$ (i.e. displacement-pressure) formulation is typically used with basis functions chosen to satisfy the Ladyzhenskaya–Babuška–Brezzi (LBB) condition (that displacement must be solved to a higher order than pressure)  \citep{sussman1987Finite,brink1996Mixed}. In this section, we will investigate the application of this technique to our osmosis-dominated hydrogel swelling problem, beginning with equilibirum problems in \cref{sec:equilibrium} before proceeding to disscussing techniques for solving the full transient system in \cref{sec:transient}.

\subsection{Uniform equilibrium problems}
\label{sec:equilibrium}
We first consider equilibrium problems for which there is no fluid flow in the gel at the steady state. Setting the time derivative and material flux $\mathbf{q}$ to zero in \cref{eq:diffeq}, we find that at equilibirum $\phi$ satisfies Laplace's equation. Assuming no porosity gradient is imposed via the boundary conditions (i.e. that all boundary conditions on $\phi$ have the form of either \labelcref{eq:phiperm} or \labelcref{eq:phiimperm} with only a single $\phi_\star$ value present), regardless of the geometry this will be solved by 
\begin{equation}
    \phi = \phi_\star \quad\text{ throughout } \Omega. \label{eq:phiuniform}
\end{equation}
In other words, the osmosis-dominated swelling model typically predicts uniform porosity at equilibrium in absence of factors such as chemcial gradients or imposed flow. (In the case of exclusively Neumann boundary conditions on $\phi$, i.e. \labelcref{eq:phiimperm}, $\phi_\star$ is not defined but $\phi$ will instead be uniformly equal to its initial value.) 

We focus in this subsection on the problem of determining the equilibrium solid deformation from \cref{eq:momeq,eq:Jcond} corresponding to a uniform equilibrium porosity \labelcref{eq:phiuniform}. This problem can be summarised as
\begin{align}
    \nabla P = \nabla\cdot\boldsymbol\sigma^\prime, \label{eq:momequniform}\\
    \boldsymbol\sigma^\prime = J^{-1}(\mathbf{F}\mathbf{F}^T-\mathbf{I}),\label{eq:neohookean}\\
     J = \frac{1-\phi_0}{1-\phi_\star} \coloneq J_\star, \label{eq:Jconst}
\end{align}
coupled to relevant boundary conditions on stress or displacement, \labelcref{eq:mech1,eq:mech2}. This problem, wherein the Jacobian $J$ is prescribed to be constant, is a close mathematical equivalent to the incompressible neo-Hookean problem wherein $J$ is prescribed to be 1, the difference being simply an effective matter of a scale factor applied to $\mathbf{F}$. 

We thus solve this problem using finite element techniques typically applied to incompressible hyperelastic problems. Treating $\phi_\star$ as a known constant, we impose the constraint \labelcref{eq:Jconst} directly in the weak form with $P$ acting as a Lagrange multiplier. We formulate the weak problem in the Lagrangian frame and seek $\boldsymbol\xi$ and $P$ such that
\begin{equation}
    \int_{\Omega_0} \mathbf{P}\colon\nabla_0\mathbf{v}+q\left(J-J_\star\right)\ dV_0-\int_{\partial\Omega_{0}} \mathbf{n}_0\cdot\mathbf{T}\cdot\mathbf{v}\ dS_0 = 0, \label{eq:solid_weak}
\end{equation}
for all admissible test functions $(\mathbf{v},q)$. Here, $\mathbf{P} = J\boldsymbol(-P\mathbf{I}+\boldsymbol\sigma^\prime)\mathbf{F}^{-T}$ is the first Piola-Kirchhoff stress tensor, $\mathbf{T}$ is the traction vector in the Lagrangian frame, $\Omega_0$ is the reference domain, $\mathbf{n}_0$ the normal to the surface $\partial\Omega_0$, and $\nabla_0$, $\mathrm{d}S_0$, $\mathrm{d}V_0$ represent differentiation and integration with respect to the reference coordinate $\mathbf{x}_0$. We discretised $\boldsymbol\xi$ and $P$ using $P_2$-$P_1$ Taylor-Hood elements, a common discretisation which has been shown to perform well in comparison to other finite element methods for incompressible hyperelasticity \citep{chamberland2010Comparison,bouklas2015Nonlinear}. The nonlinear equations are implemented in FEniCSx \citep{alnaes2014Unified,scroggs2022Basix,scroggs2022Construction,baratta2023DOLFINx} and solved using a Newton-Raphson method with backtracking linesearch, automatic differentiation, and linear systems solved directly using MUMPS \citep{amestoy2001Fully}.

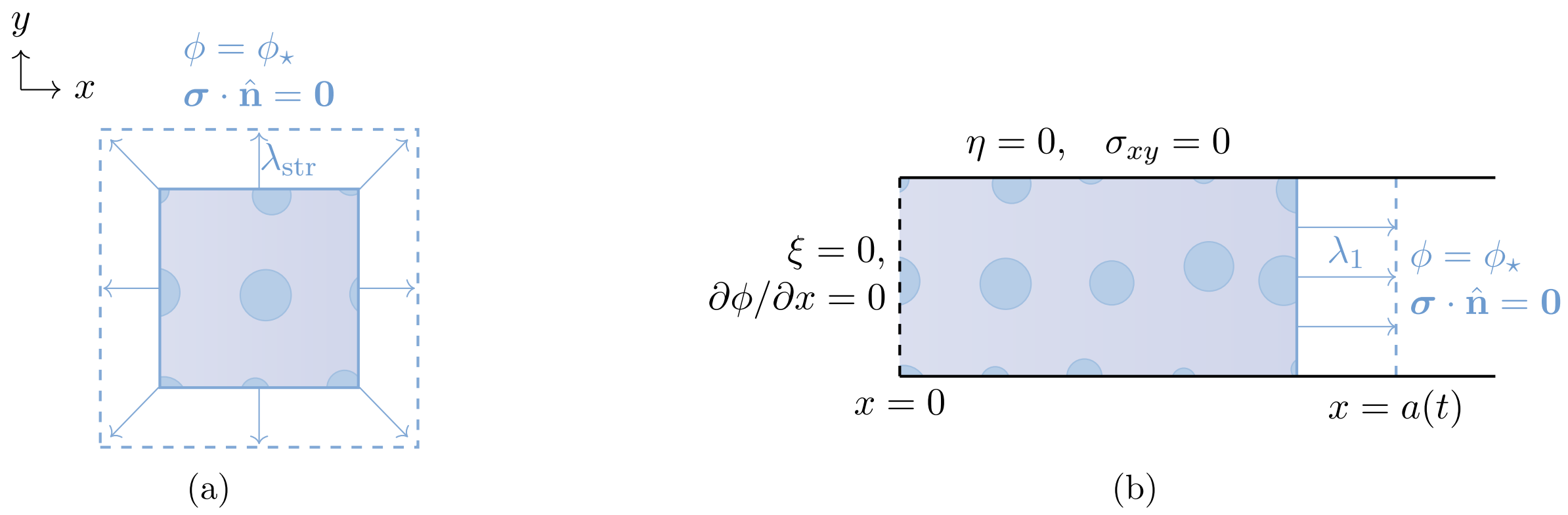
\begin{figure}
    \centering
    \begin{subfigure}{0.33\textwidth}
        \centering
                \tikzsetnextfilename{porous_square}
    \begin{tikzpicture}

                \begin{scope}
                \clip[preaction={shading=radial, left color=gel!80, right color=gel}, 
        ] (0,0) -- (2,0) -- (2,2) -- (0,2) -- cycle;

        \foreach \i in {0,...,4}{
            \foreach \j in {-2,...,2}{
            \pgfmathsetmacro{\x}{\i+rand*0.15};
            \pgfmathsetmacro{\y}{\j+rand*0.15};
            \pgfmathsetmacro{\r}{0.2+0.1*rand};
            \fill[outline!35] (\x,\y) circle (\r cm);
            \draw[outline!45] (\x,\y) circle (\r cm);
        }
        }
        \end{scope}

         \draw[outline!60,thick] (0,0) -- (2,0) -- (2,2) -- (0,2) -- cycle;
         \draw[outline!60,thick,dashed] (-0.6,-0.6) -- (2.6,-0.6) -- (2.6,2.6) -- (-0.6,2.6) -- cycle;
         \foreach \i in {0,...,3}{
            \pgfmathsetmacro{\th}{45+\i*90};
            \pgfmathsetmacro{\r}{2.12};
            \draw[->,outline!60] ({1+1.43*cos(\th)},{1+1.41*sin(\th)}) -- ({1+\r*cos(\th)},{1+\r*sin(\th)});
         }
                  \foreach \i in {0,...,3}{
            \pgfmathsetmacro{\th}{\i*90};
            \pgfmathsetmacro{\r}{1.57};
            \draw[->,outline!60] ({1+cos(\th)},{1+sin(\th)}) -- ({1+\r*cos(\th)},{1+\r*sin(\th)});
         }
         \node[anchor=south,outline!70,align=left] at (1,2.7) {$\phi=\phi_\star$ \\ $\boldsymbol\sigma\cdot\hat{\mathbf{n}} = \mathbf{0}$};

         \node[outline!70] at (1.3,2.3) {$\lambda_\text{str}$};
\begin{scope}[shift={(-0.2,4)}]
         \draw[->] (-1.2,-1) -- (-0.8,-1) node[anchor=west] {$x$};
         \draw[->] (-1.2,-1) -- (-1.2,-0.6) node[anchor=south] {$y$}; 
\end{scope}
    \end{tikzpicture}
    \caption{}
\end{subfigure}
\begin{subfigure}{0.66\textwidth}
    \centering
            \tikzsetnextfilename{porous_bar}
       \begin{tikzpicture}

                \begin{scope}
                \clip[preaction={shading=radial, left color=gel!80, right color=gel}, 
        ] (0,0) -- (4,0) -- (4,2) -- (0,2) -- cycle;

        \foreach \i in {0,...,4}{
            \foreach \j in {-2,...,2}{
            \pgfmathsetmacro{\x}{\i+rand*0.15};
            \pgfmathsetmacro{\y}{\j+rand*0.15};
            \pgfmathsetmacro{\r}{0.2+0.1*rand};
            \fill[outline!35] (\x,\y) circle (\r cm);
            \draw[outline!45] (\x,\y) circle (\r cm);
        }
        }
        \end{scope}

         \draw[outline!60,thick] (4,0) -- (4,2);
         \draw[outline!60,thick,dashed] (5,0) -- (5,2);
         \draw[thick] (0,0) -- (6,0); 
         \draw[thick] (0,2) -- (6,2);
         \draw[thick,dashed] (0,0) -- (0,2);
         \node[anchor=north] at (0,0) {$x=0$};
         \node[anchor=north] at (5,0) {$x = a(t)$};
         \foreach \i in {1,...,3}{
            \draw[->,outline!60] (4,{\i*0.5}) -- (5,{\i*0.5});
         }
         \node[anchor=west,outline!70,align=left] at (5,1) {$\phi=\phi_\star$ \\ $\boldsymbol\sigma\cdot\hat{\mathbf{n}} = \mathbf{0}$};

         \node[outline!70] at (4.5,1.25) {$\lambda_1$};
         \node[anchor=south,align=center] at (2,2) {$\eta = 0$,\quad $\sigma_{xy} = 0$};
         \node[anchor=east,align=right] at (0,1) {$\xi = 0$,\\ $\partial\phi/\partial x = 0$};
\node at (-0,-0.6) {};
    \end{tikzpicture}

    \caption{}
\end{subfigure}
    \caption{A diagram showing two problems which we use as numerical benchmarks. In (a), a hydrogel square swells isotropically to equilibrium stretch $\lambda_\text{str}$. We impose $\phi = \phi_\star$ and zero normal stress on the free surface. In (b), a hydrogel bar between two parallel rigid walls swells uniaxially to equilibrium stretch $\lambda_1$ in the $x$ direction. We again impose $\phi = \phi_\star$ and zero normal stress on the free surface, alongside no normal displacement and no tangential stress on the walls with slip, and symmetry conditions at $x=0$.}
    \label{fig:benchmarks_diagram}
\end{figure}

\begin{figure}
    \centering
            \tikzsetnextfilename{porous_wedge}
    \begin{tikzpicture}

                \begin{scope}
                \clip[preaction={shading=radial, left color=gel!80, right color=gel}, 
        ] (0,0) -- (3.464,2) arc[start angle=30,end angle=-30,radius=4] -- cycle;

        \foreach \i in {0,...,4}{
            \foreach \j in {-2,...,2}{
            \pgfmathsetmacro{\x}{\i+rand*0.15};
            \pgfmathsetmacro{\y}{\j+rand*0.15};
            \pgfmathsetmacro{\r}{0.2+0.1*rand};
            \fill[outline!35] (\x,\y) circle (\r cm);
            \draw[outline!45] (\x,\y) circle (\r cm);
        }
        }

            \draw[black!80] (0.866,0.5) arc[start angle=30, end angle=-30, radius=1];
            \node[black!80] at (0.6,0) {$2\beta$};

        \end{scope}
                \draw[thick] (0,0) -- (3.464,2);
                \draw[thick] (0,0) -- (3.464,-2);
        \foreach \i in {0,...,3}{
            \draw ({(\i+0.5)*0.866},{(\i+0.5)*0.5}) -- ({(\i+0.5)*0.866+0.3},{(\i+0.5)*0.5+0.5});
            \draw ({(\i+0.5)*0.866},{-(\i+0.5)*0.5}) -- ({(\i+0.5)*0.866+0.3},{-(\i+0.5)*0.5-0.5});
        }
         \draw[outline!60,thick] (3.464,2) arc[start angle=30,end angle=-30,radius=4];
         \draw[outline!60,thick,dashed] (3.464,2) to[out=335,in=25] (3.464,-2);
         \foreach \i in {0,...,4}{
            \pgfmathsetmacro{\th}{-20+\i*10};
            \pgfmathsetmacro{\r}{4.5-0.01*abs(\th)};
            \draw[->,outline!60] ({4*cos(\th)},{4*sin(\th)}) -- ({\r*cos(\th)},{\r*sin(\th)});
         }
         \node[anchor=west,outline!70,align=left] at (4.6,0) {$\phi=\phi_\star$ \\ $\boldsymbol\sigma\cdot\hat{\mathbf{n}} = \mathbf{0}$};
         \node at (1.5,-1.7) {$\boldsymbol\xi = \mathbf{0}$};
         \node[anchor=north east] at (0,0) {$O$};
        \node[anchor=south west] at (3.464,2) {$r=1$};

            \begin{scope}[shift={(-0.5,2)}]
    \def\polarAngle{40}    
    \def\radius{1.3}        

    \coordinate (rvec) at ({\radius*cos(\polarAngle)}, {\radius*sin(\polarAngle)});
    \draw[->, thick] (0,0) -- (rvec) node[anchor=south west] {$r$};

    \draw[->, thick] 
      ([shift={(0.7,0)}]0,0) arc[start angle=0, end angle=\polarAngle, radius=0.7];
    \node at (0.9,0.3) {$\theta$};
    \draw[dashed] (0,0) -- (\radius,0);
            \end{scope}
    \end{tikzpicture}
    \caption{A diagram showing a hydrogel wedge adhered to rigid walls at $\theta = \pm \beta$, with a free surface initially at $r=1$. We impose zero displacement at the walls and $\phi=\phi_\star$ and no normal stress at the free surface.}
    \label{fig:wedge_diagram}
\end{figure}
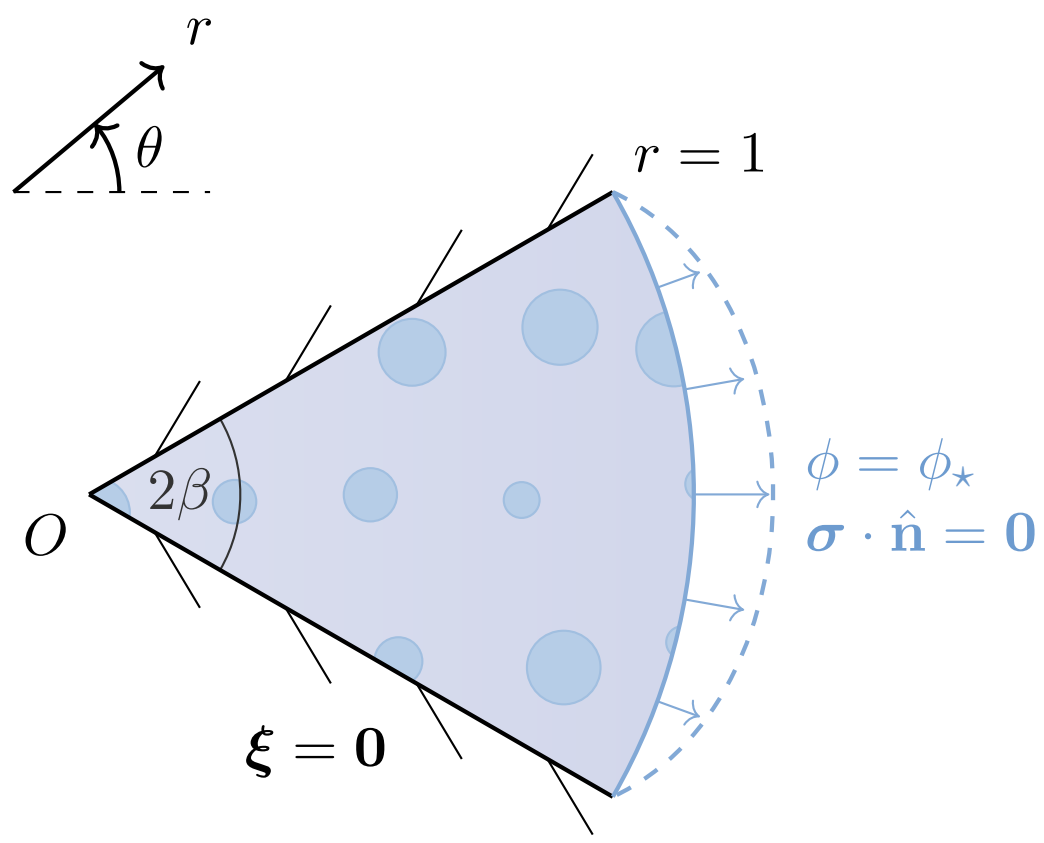

\begin{figure}
    \centering
         \begin{subfigure}{0.49\textwidth}
            \includegraphics[width=\textwidth]{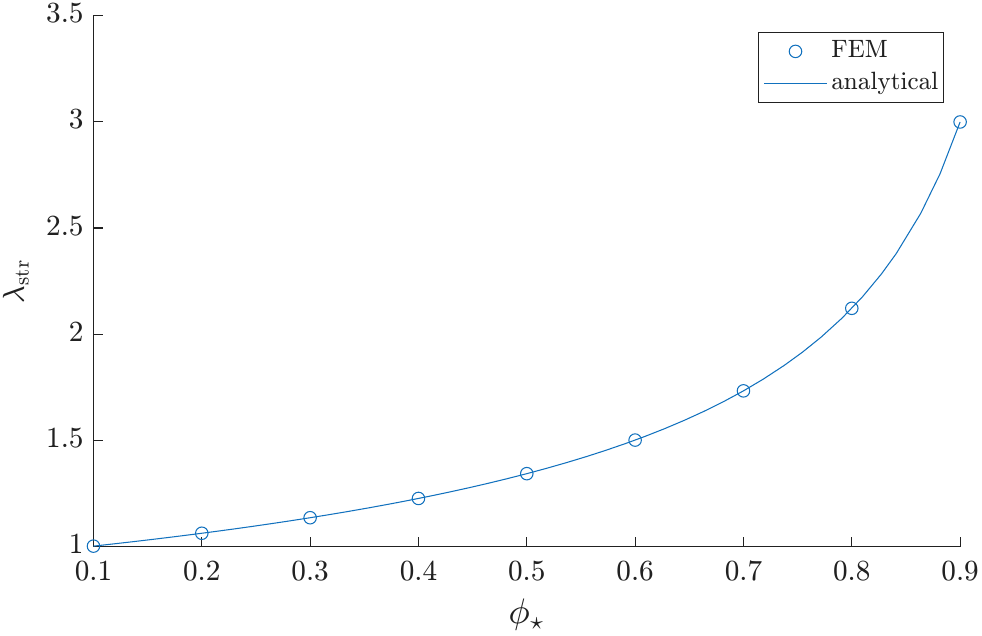}
            \caption{}
            \label{fig:square}
        \end{subfigure}
        \begin{subfigure}{0.49\textwidth}
            \includegraphics[width=\textwidth]{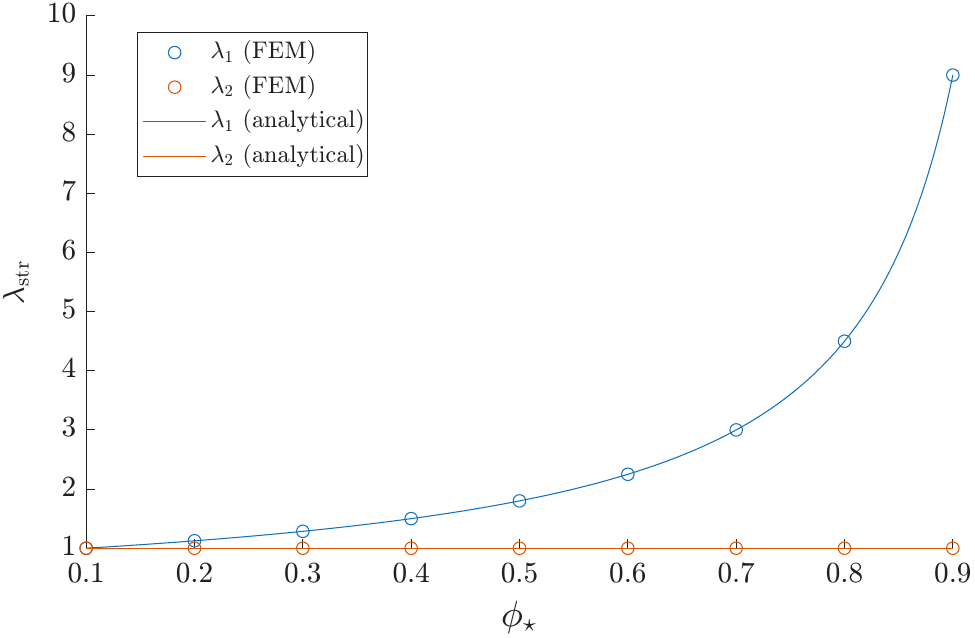}
            \caption{}
            \label{fig:uniaxial}
        \end{subfigure} 
        \caption{Numerical benchmarks for the equilibrium finite element method. In (a), the isotropic stretch for a free-swelling square hydrogel compared to the analytical value; in (b) the horizontal ($\lambda_1$) and vertical ($\lambda_2$) stretches for the uniaxial swelling of hydrogel bar compared with their analytical values.}
        \label{fig:benchmark}
\end{figure}

As benchmarks for the finite element method we consider two problems for which analytical solutions can be found. For simplicitly, we focused in this paper on two-dimensional problems (by which we mean plane-strain problems for which we assume symmetry in the third dimension). The two problems are illustrated in \cref{fig:benchmarks_diagram}. Firstly, we consider a free-swelling hydrogel square (i.e. the cross-section of a cuboid of infinite length). We took the square to be of unit length and imposed traction free boundary conditions on each edge. The analytical solution to this problem for given $\phi_\star$ is isotropic expansion with stretch $\lambda_\mathrm{str} = \sqrt{J_\star}$.  Secondly, we consider the uniaxial extension of a hydrogel bar. A rectangular hydrogel is held fixed between two frictionless rigid walls; thus, a no-slip condition is applied to the normal displacement but a stress-free condition in the tangential direction on these surfaces. In both cases, we applied the finite element method described above with varying $\phi_\star$ values and computed the relevant stretches at approximately evenly spaced sample points in the domain. The results are shown in \cref{fig:benchmark} and show very close agreement. In both cases, the domain was discretized using a uniform $0.1\times 0.1$ grid, with each square cell split along the diagonal to form two triangular elements.

\begin{figure}
    \centering
        \begin{tabular}{c c c c}
             $\phi_\star = 0.4$ & $\phi_\star = 0.6$ & $\phi_\star = 0.8$ & \\
         \begin{minipage}{0.3\textwidth}
            \includegraphics[width=\textwidth,trim=0.9cm 0.8cm 1.8cm 0,clip]{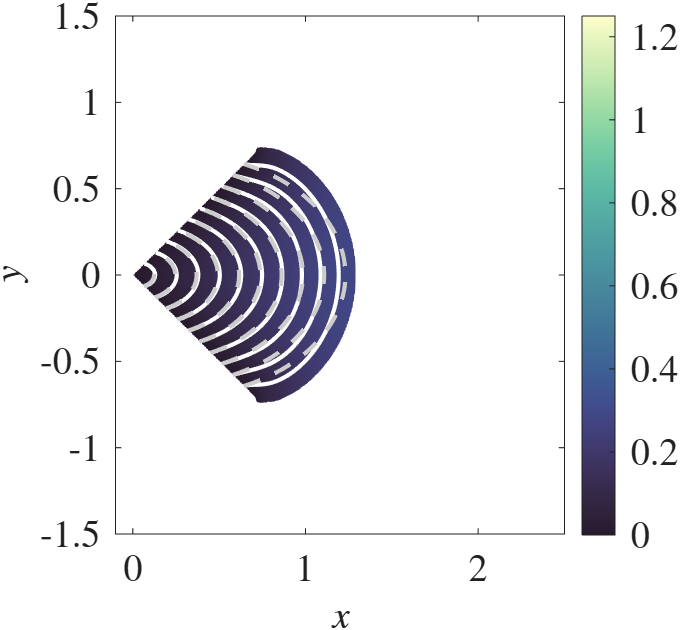}
        \end{minipage}&
        \begin{minipage}{0.27\textwidth}
            \includegraphics[width=\textwidth,trim=1.8cm 0.8cm 1.8cm 0,clip]{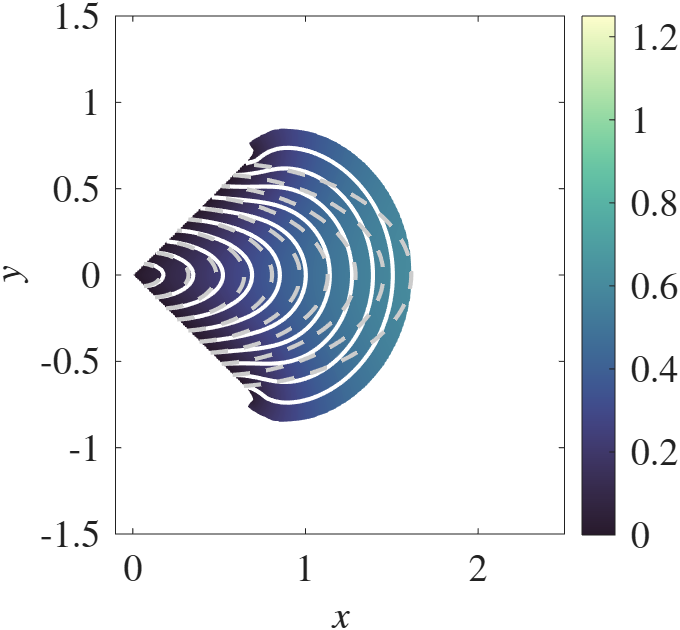}
        \end{minipage} &
                \begin{minipage}{0.33\textwidth}
            \includegraphics[width=\textwidth,trim=1.8cm 0.8cm 0 0,clip]{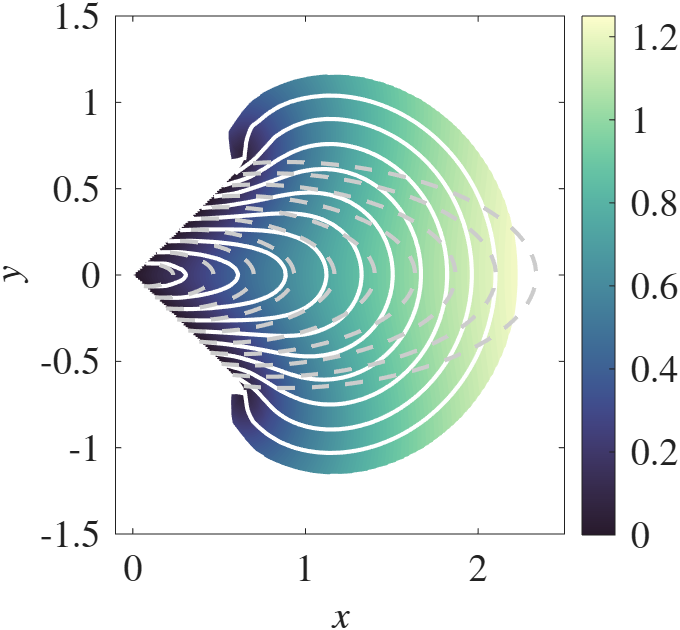}
        \end{minipage} &
                \begin{minipage}{0.03\textwidth}
        $|\hat{\boldsymbol\xi}|$
    \end{minipage}
\end{tabular}
        \caption{Equilibrium swelling of a fully nonlinear adhered hydrogel wedge computed by finite element methods. The contours in white show the deformation of material curves which were circular arcs in the reference configuration, i.e. $|\mathbf{x}_0| = $const. The grey dashed lines show the equivalent contours for the analytic linear solution displayed in \cref{fig:wedge}. Initial porosity was $\phi_0 = 0.1$.}
        \label{fig:femwedge}
\end{figure}
In \cref{fig:femwedge} we return to the adhered wedge geometry of \cref{sec:wedge} but consider now the fully nonlinear problem with a free surface present; see \cref{fig:wedge_diagram}. We have plotted the equilibrium displacement profile of the swollen gel for $\phi_\star = 0.4, 0.6, 0.8$, which for a reference porosity $\phi_0 = 0.1$ corresponds to $J_\star = 1.5, 3, 4.5$ respectively. In the results, we have overlayed the finite element solution with the analytical linear solution \labelcref{eq:wedge-soln}. We see that the two solutions agree closely near the adhered boundaries in all three cases and for $\phi_\star = 0.4$ (and, to a lesser extent $\phi_\star = 0.6$) they agree closely for the bulk of the domain; this is perhaps surprising since a change in $\phi$ from 0.1 to 0.4 or 0.6 might not be considered especially `small'. Note that as well as the linear-nonlinear distinction between the two solutions, the linear solution was for an infinite wedge, so even for small displacements we expect to two solutions to differ near the free surface. For $\phi_\star = 0.8$, we see more substantial disagreement between the two solutions, and near the free surface the gel has `bulged' from the sides of the wedge. The wedge domain was discretized using Gmsh \citep{geuzaine2009Gmsh} into a uniform triangular mesh with characteristic element size 0.05.
\subsection{Transient problems}
\label{sec:transient}
As discussed in the opening of this section, the transient nonlinear problem involves nontrivial coupling between fluid transport and solid deformation. However, the assumption of small $G$ provides simplifications in the form of omitting elastic stress in the diffusion equation \labelcref{eq:diffeq}, the formulation of boundary conditions in terms of $\phi$, and the reduction of the solid problem to a form closely related to incompressible hyperelasticity but with modified constraint \labelcref{eq:Jcond}. 

The natural split of our osmosis-dominated system into a diffusion problem and a constrained hyperelastic problem leads us to postulate that a staggered finite element procedure may be an effective numerical method for its solution. By staggered, we mean a procedure wherein, on each time step:
\begin{enumerate}
    \item \hspace{0.1cm} the porosity is found assuming the solid deformation of the previous time step;
    \item \hspace{0.1cm} the solid deformation is found given the porosity computed in step 1.
\end{enumerate}
It is possible also to iterate the two steps within a single time-step to achieve better accuracy. 

As discussed in \cref{sec:introduction}, there have been few studies investigating the application of staggered techniques to hydrogel swelling problems, with most authors opting for a monolithic approach \citep{bouklas2015Nonlinear,zhang2009Finite,liu2016Transient}. However, a review by \citet{urrea-quintero2024Comparative} found that staggered methods achieved comparable accuracy and stability to monolithic methods for a particular constitutive model (called `Constitutive Model II' in the paper, and based on that of \citet{chester2010Coupled}). This model had the key distinguishing features that it (a) imposed an incompressible criterion similar to \cref{eq:Jcond} via a pressure-type Lagrange multiplier and (b) focused on simple geometries for which that pressure could be expressed as a direct function of $\phi$. Since these features are both shared by our osmosis-dominated regime, this provides evidence that staggered approaches are effective for finite element formulations with the same mathematical structure as that which we propose here. In the review, they restricted use of this method to simple geometries on the basis that $p$ is not normally easily found as a straightforward function of $\phi$. In our osmosis-dominated regime, however, $p \approx -\Pi(\phi)$, eliminating this difficulty and thus broadening the range of problems to which this method is potentially applicable.

We derive the weak form of the system \labelcref{eq:diffeq,eq:momeq,eq:Jcond} in the Lagrangian frame to obtain
\begin{multline}
    \int_{\Omega_0}\ \psi J(\phi)\left(\frac{\mathrm{d}\phi}{\mathrm{d}t}+D(\phi)\nabla_0\phi\cdot\mathbf{F}^{-1}\mathbf{F}^{-T}\cdot\nabla_0\phi\right) \\ +D(\phi)J(\phi)\nabla_0\phi \cdot\mathbf{F}^{-1}\mathbf{F}^{-T}\cdot\nabla_0\psi\ \mathrm{d}V_0, \label{eq:diff_weak}
\end{multline}
for all admissible test functions $\psi$, where boundary terms vanish on the assumption that $\phi$ satisfies either \cref{eq:phiimperm} or \cref{eq:phiperm} on all boundaries. The weak form for the corresponding solid mechanics problem is the same as \labelcref{eq:solid_weak} but we now enforce $J = J(\phi) = (1-\phi_0)/(1-\phi)$ rather than $J = J_\star$.

Let the superscript $n$ represent the computed value of a quantity at the $n$th time step. We apply a backwards Euler discretization in time with time step $\Delta t$ to \cref{eq:diff_weak}, and on each time step approximate the displacement by that of the previous time step,
    \begin{multline}
    \int_{\Omega_0}\ \psi J(\phi^{n+1})\left(\frac{\phi^{n+1}-\phi^n}{\Delta t}+D(\phi^{n+1})\nabla_0\phi^{n+1}\cdot\left(\mathbf{F}^{-1}\right)^n\left(\mathbf{F}^{-T}\right)^n\cdot\nabla_0\phi^{n+1}\right) \\ +D(\phi^{n+1})J(\phi^{n+1})\nabla_0\phi^{n+1} \cdot\left(\mathbf{F}^{-1}\right)^n\left(\mathbf{F}^{-T}\right)^n\cdot\nabla_0\psi\ \mathrm{d}V_0. \label{eq:diff_disc}
\end{multline}
On time step $n+1$, we solve for $\phi^{n+1}$ from \cref{eq:diff_disc} with $\phi^n$ and $\mathbf{F}^n$ known from the previous step. We discretize $\phi$ using $P_1$ Lagrange elements and solve the resulting nonlinear equation by a Newton-Raphson method with backtracking linesearch and automatic differentiation, again using MUMPS as the linear solver. We then solve the solid mechanics problem using the updated porosity value,
\begin{equation}
    \int_{\Omega_0} \mathbf{P}^{n+1}\colon\nabla_0\mathbf{v}+q\left(J-J(\phi^{n+1})\right)\ dV_0-\int_{\partial\Omega_{0}} \mathbf{n}_0\cdot\mathbf{T}\cdot\mathbf{v}\ dS_0 = 0. \label{eq:solid_disc}
\end{equation}
We solve \cref{eq:solid_disc} using the same scheme as outlined in \cref{sec:equilibrium} for equilibrium problems. We then proceed to time step $n+2$. Note that, as mentioned earlier in this section, for improved accuracy we may iterate these two steps to convergence within a single time step before proceeding to the next.

\begin{figure}
    \centering
    \includegraphics[width=0.8\textwidth]{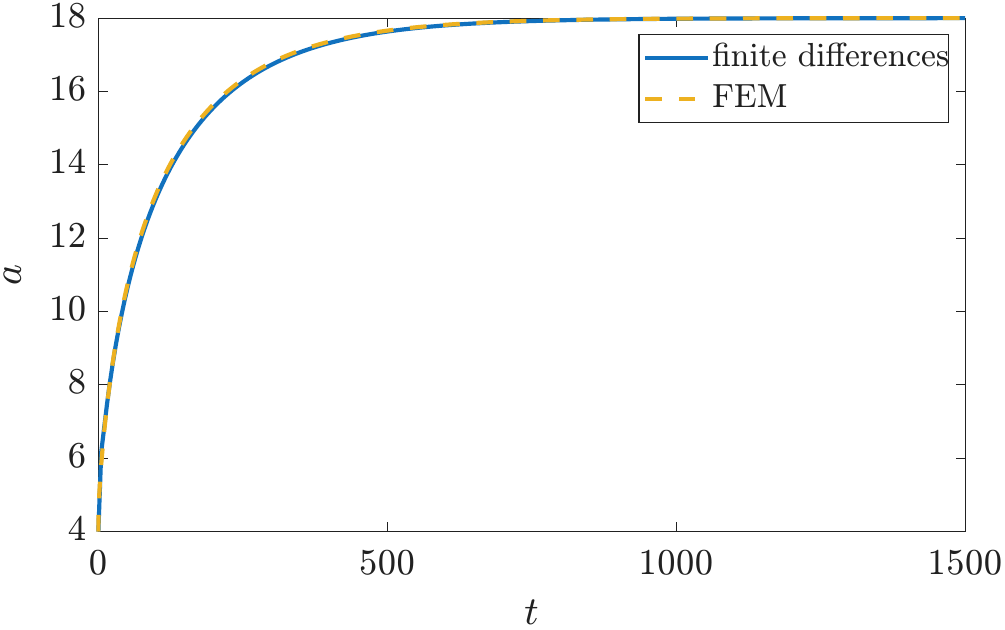}
    \caption{A comparison of the free surface position $a(t)$ as computed by both FEM and finite differences. For 2D FEM, we used a $2\times 4$ rectangular domain, discretized into a uniform $0.1\times 0.1$ grid, with each square cell split along the diagonal to form two triangular elements. The time step was $\Delta t = 5$. For finite differences, we treated the problem as 1D and used a spatial step $\Delta y = 0.01$ and time step $\Delta t = 0.1$.}
    \label{fig:transient_uniaxial}
\end{figure}

As a simple check for this scheme, we again consider the problem of uniaxial swelling but now consider the transient problem. Since this problem is 1D, it can be solved by finite differences since the horizontal displacement $\xi$ is related to $\phi$ by $J(\phi) = 1-\partial\xi/\partial x$. This relationship makes it possible to write the Lagrangian diffusion equation in terms of only $\phi$ and therefore we require no solid mechanics solution step. For the finite difference method, we use central differences in space and a backwards Euler discretization in time, with the resulting nonlinear equations solved by MATLAB's fsolve function. In \cref{fig:transient_uniaxial}, we plot the position of the free surface $x = a(t)$ as computed by both methods, and they agree very closely. 

\begin{figure}
    \centering
        \begin{tabular}{c c c c}
             $t = 0.25$ & $t = 1$ & $t = 5$ & \\
         \begin{minipage}{0.3\textwidth}
            \includegraphics[width=\textwidth,trim=0.9cm 0.8cm 1.8cm 0,clip]{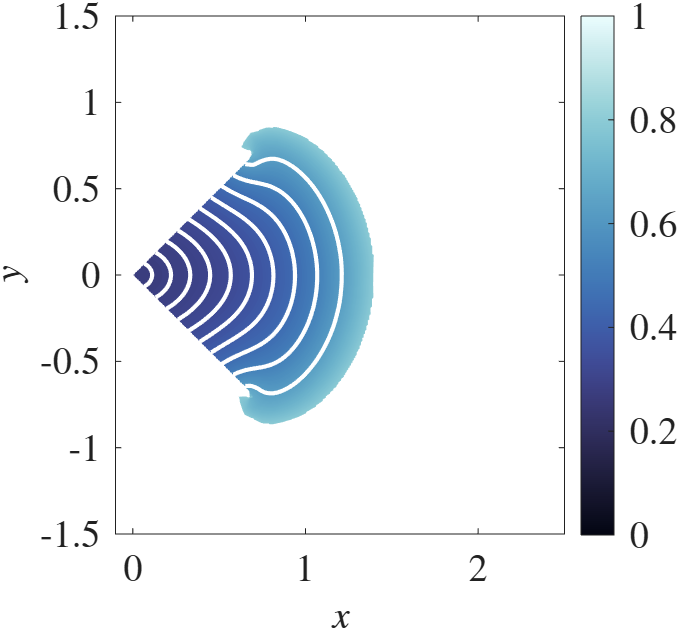}
        \end{minipage}&
        \begin{minipage}{0.27\textwidth}
            \includegraphics[width=\textwidth,trim=1.8cm 0.8cm 1.8cm 0,clip]{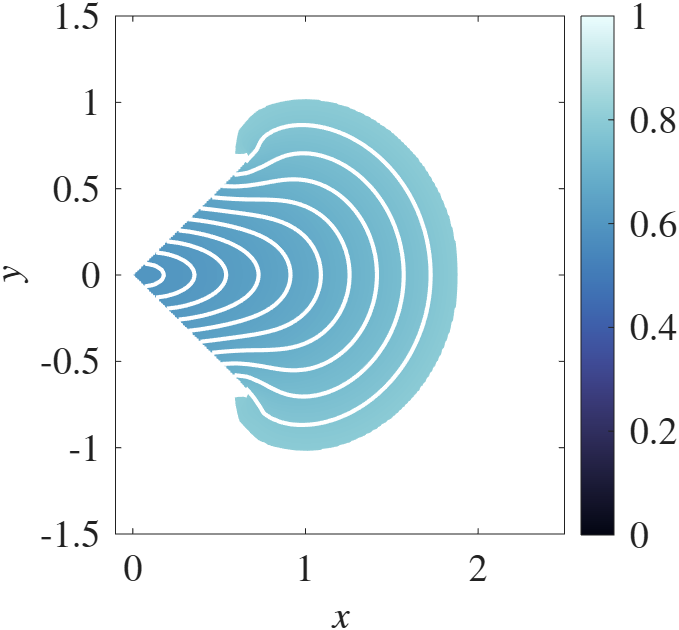}
        \end{minipage} &
                \begin{minipage}{0.33\textwidth}
            \includegraphics[width=\textwidth,trim=1.8cm 0.8cm 0 0,clip]{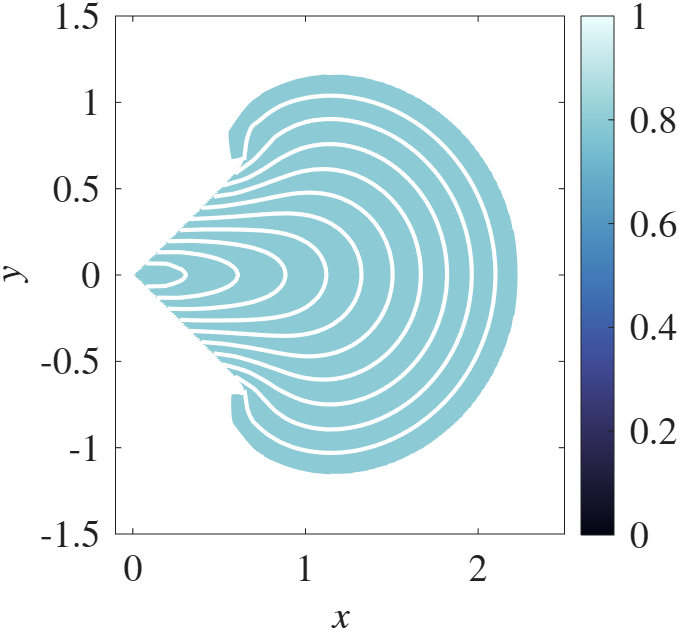}
        \end{minipage} &
                \begin{minipage}{0.03\textwidth}
        $\phi$
    \end{minipage}
\end{tabular}
        \caption{Transient swelling of a fully nonlinear adhered hydrogel wedge computed by a finite element method. The contours in white show the deformation of material curves which were circular arcs in the reference configuration, i.e. $|\mathbf{x}_0| = $const. Initial porosity was $\phi_0 = 0.1$ and the fully swollen value was $\phi_\star = 0.8$.}
        \label{fig:transient_wedge}
\end{figure}

Finally, we solve for the transient dynamics of the adhered hydrogel wedge problem shown in \cref{fig:wedge_diagram}. Using the same mesh as for the equivalent equilibrium problem (produced by Gmsh with characteristic length 0.05) and a time step $\Delta t = 0.25$ we obtain the results shown in \cref{fig:transient_wedge}. We see the wedge converged to the same equilibirum result showin in the third panel of \cref{fig:femwedge}, while at earlier times there are larger gradients in porosity present.

\section{Conclusion}
\label{sec:conclusion}
In this paper, we have derived an asymptotic regime for hydrogel swelling which we term the `osmosis-dominated' regime. This regime is based on the principle that the elastic contributions to hydrogel free energy are much smaller in characteristic scale than the osmotic contributions, resulting in their ratio $G$ being a small parameter. This idea has support from both theory and experiment \citep{cai2011mechanics,bertrand2016dynamics,correas2026Hydrogel}, to the extent that we would argue it could be considered a defining feature of hydrogels (in the sense that swelling is often considered a defining property, and hydrogels with large $G$ would not swell significantly). 

In the limit of small deformation, this reduces the equations of hydrogel mechanics to a hierarchy of three linear PDE problems (diffusion, Poisson and Stokes) which may be solved sequentially using the relevant classical tehcniques developed for them. We use hierarchy here in the sense that the coupling is only one-way; the porosity may be solved for independently of displacement, and displacement made be solved for given the porosity. This is a significant simplification from the fully coupled system and offers a straightforward strategy by which to solve the equations governing hydrogel swelling in a wide variety of geometries using textbook methods.

In the finite deformation case, the coupling remains two-way but is weakened; furthermore, the ability to formulate boundary conditions directly in terms of porosity rather than pressure or chemical potential enables fintie element formulations using porosity directly as an independent variable. The result of this is that uniform equilibrium problems are reduced to a problem directly analogous to incompressible hyperelasticity, and as such one can easily adapt numerical schemes developed for incompressible hyperleastic materials to hydrogels. We note that although we have demonstrated proof of concept for this idea, and have been able to reproduce analytical results for simple geometries, the criterion imposed on the Jacobian in our hydrogel model is distinct from the standard $J=1$ incompressible criterion and therefore a more thorough study would be required to establish the accuracy and stability of these techniques across geometries.

For transient problems, we have proposed that the equations in our regime make a staggered finite element approach natural. Staggered approaches are common in poroelasticity but uncommon in hydrogel mechanics \citep{zienkiewicz1988Unconditionally,urrea-quintero2024Comparative}. We have shown that our scheme produces results which match well with those of standard finite difference methods for a simple uniaxial geometry. For the adhered wedge problem, we have no benchmark to compare to, but the scheme produces physically reasonable results with the expected equilibrium state. We note that in this work we used a very straightforward splitting technique with no iterations within each time-step, and no adaptive meshing or adaptive time-stepping; one would expect improved accuracy with these measures in place. This would be an interesting subject for further work.

The techniques we outline in this paper have broad applicability to hydrogel swelling problems in both small and large deformation cases, and in future work could be applied to a wide range of geometries. This has great potential for developing simplified models of hydrogel behaviour for a range of applications, including drug delivery, wound healing, and soft robotics. An interesting avenue for further work  would be the development of boundary element methods (BEM) for hydrogels; the linear equations derived in \cref{sec:linear} essentially result in solving a relevant Stokes flow problem, and as such the standard numerical techniques which have been very well-developed for Stokes flows over the decades \citep{pozrikidis2002Practical} can in principle be applied here. This would in principle provide a general method by which to solve the linear hydrogel swelling equations in arbitrary geometry. \\ \\

\small \noindent \textbf{Acknowledgements.} The authors thank Joseph Webber for helpful discussions and comments on a first draft of the manuscript. \\ \\

\noindent \textbf{AI usage declaration.} All conceptualization, analysis, and writing involved in the production of this article was undertaken by the authors. Generative AI tools were consulted during software development and literature review but did not generate the final code, nor any text or references used in the article. The authors take full responsibility for the content of this article. \\ \\

\noindent \textbf{Funding.} This work was supported by the Leverhulme Trust Research Leadership Award ‘Shape-Transforming Active Microfluidics’ (RL-2019-014) to TDMJ at the University of Warwick, on which EMJ is employed as a postdoctoral research fellow. \\ \\

\noindent \textbf{Declaration of interests.} The authors report no conflict of interest. \\ \\

\noindent \textbf{Data availability statement.} The code used in this study can be shared on request. 

\bibliographystyle{plainnat}
\bibliography{smallG}
\end{document}